\documentclass[aps,prx,twocolumn,footinbib,floatfix,superscriptaddress]{revtex4-2}
\usepackage{graphicx}
\usepackage{indentfirst}
\usepackage{physics}
\usepackage{braket}
\usepackage{float}
\usepackage{amsmath}
\usepackage{epstopdf}
\usepackage{footnote}
\usepackage{CJK}
\usepackage{esint}
\usepackage{color}
\usepackage[T1]{fontenc}
\usepackage{subfigure}
\usepackage{amsfonts}
\usepackage{footmisc}
\usepackage{scrextend}
\usepackage{multirow}
\usepackage[hyperfootnotes=false]{hyperref}
\usepackage[acronym]{glossaries}
\usepackage[english]{babel}
\usepackage{url}
\usepackage{bm}
\usepackage{hyperref}
\usepackage{mathtools}
\usepackage{svg}
\definecolor{darkblue}{rgb}{0,0,0.5}
\hypersetup{
colorlinks=true,
linkcolor=black,
filecolor=blue,
citecolor=darkblue,  
urlcolor=black,
}

\newcommand{\bp}{\boldsymbol \rho}

\newcommand*\diff{\mathop{}\!\mathrm{d}}

\newcommand{\calC}{{\cal C}}

\newcommand{\calE}{{\cal E}}

\newcommand{\1}{^{(1)}}

\newcommand{\bI}{\boldsymbol I}
\newcommand{\bZ}{\boldsymbol Z}

\newcommand{\bx}{\boldsymbol x}

\newcommand{\state}[1]{\ketbra{#1}{#1}}

\def\be{\begin{equation}}
\def\ee{\end{equation}}
\def\ba{\begin{eqnarray}}
\def\ea{\end{eqnarray}}
\def\bal{\begin{equation}\begin{aligned}}
\def\eal{\end{aligned}\end{equation}}

\def\bp{\begin{pmatrix}}
\def\ep{\end{pmatrix}}

\begin{document}

\title{
Entangling remote microwave quantum computers with hybrid entanglement swap and variational distillation
}

\author{Bingzhi Zhang}
\thanks{These two authors contributed equally.}
\affiliation{
Department of Physics, University of Arizona, Tucson, AZ 85721, USA
}
\affiliation{
Department of Electrical and Computer Engineering, University of Arizona, Tucson, Arizona 85721, USA
}
\author{Jing Wu}
\thanks{These two authors contributed equally.}
\affiliation{
James C. Wyant College of Optical Sciences, University of Arizona, Tucson, AZ 85721, USA
}
\affiliation{
Department of Electrical and Computer Engineering, University of Arizona, Tucson, Arizona 85721, USA
}

\author{Linran Fan}
\affiliation{
James C. Wyant College of Optical Sciences, University of Arizona, Tucson, AZ 85721, USA
}

\author{Quntao Zhuang}
\email{zhuangquntao@email.arizona.edu}
\affiliation{
Department of Electrical and Computer Engineering, University of Arizona, Tucson, Arizona 85721, USA
}
\affiliation{
James C. Wyant College of Optical Sciences, University of Arizona, Tucson, AZ 85721, USA
}

\begin{abstract}
Superconducting microwave circuits with Josephson junctions are a major platform for quantum computing.
To unleash their full capabilities, the cooperative operation of multiple microwave superconducting circuits is required. Therefore, designing an efficient protocol to distribute microwave entanglement remotely becomes a crucial open problem.
Here, we propose a continuous-variable entanglement-swap approach based on optical-microwave entanglement generation, which can boost the ultimate rate by two orders of magnitude at state-of-the-art parameter region, compared with traditional approaches. 
We further empower the protocol with a hybrid variational entanglement distillation component to provide huge advantage in the infidelity-versus-success-probability trade-off. Our protocol can be realized with near-term device performance, and is robust against non-perfections such as optical loss and noise. Therefore, our work provides a practical method to realize efficient quantum links for superconducting microwave quantum computers.

\end{abstract}

\date{\today}
\maketitle

Empowered by the law of quantum mechanics, quantum computers have the potential of speeding up the solution of various classically hard problems~\cite{Shor_1997,harrow2009,grover1996fast,bloch2012quantum}.
Among the candidate platforms for quantum computing, superconducting circuits with Josephson junctions stand out with high scalability and strong single-photon nonlinearity, as exemplified by recent demonstrations~\cite{arute2019quantum,wu2021strong,campagne2020quantum,eickbusch2021fast}. 
Just like their classical counterparts, quantum computers need to be connected to complete advanced computing and networking tasks. However, different from the classical case, the connection between quantum computers requires entanglement---quantum correlation beyond classical physics~\cite{kimble2008quantum,wehner2018quantum,kozlowski2019towards}.

Although direct microwave links can be used at short distances for proof-of-principle demonstrations~\cite{zhong2021deterministic}, the connection between superconducting quantum computing circuits over long distances is best implemented with optical photons via microwave-optical transduction~\cite{han2021microwave,andrews2014bidirectional,bochmann2013nanomechanical,fan2018superconducting,xu2020bidirectional,hisatomi2016bidirectional,williamson2014magneto,bartholomew2020chip}. Only optical photons can maintain the quantum coherence with the low dissipation and decoherence rate at room temperature. In spite of significant improvement, it is still challenging to realize high-performance microwave-optical transduction with near-unity efficiency and near-zero noise. Then a critical question is---what is the most efficient way to connect superconducting microwave quantum computers given the non-ideal performance of microwave-optical transduction. We have recently proved that continuous-variable (CV) teleportation can provide a much higher capacity for microwave-optical transduction, compared with direct conversion schemes~\cite{wu2021deterministic}. 
To leverage the capacity advantage, we still need to address critical questions including (i) how the entanglement rate between two remote superconducting microwave circuits can be improved by the higher microwave-optical transduction capacity, (ii) how capacity improvement can translate into a practical high-rate (in general CV) protocol, (iii) how noisy CV entanglement can be converted into high-fidelity discrete-variable (DV) entanglement between superconducting microwave quantum computers.

\begin{figure*}
    \centering
    \includegraphics[width=0.95\textwidth]{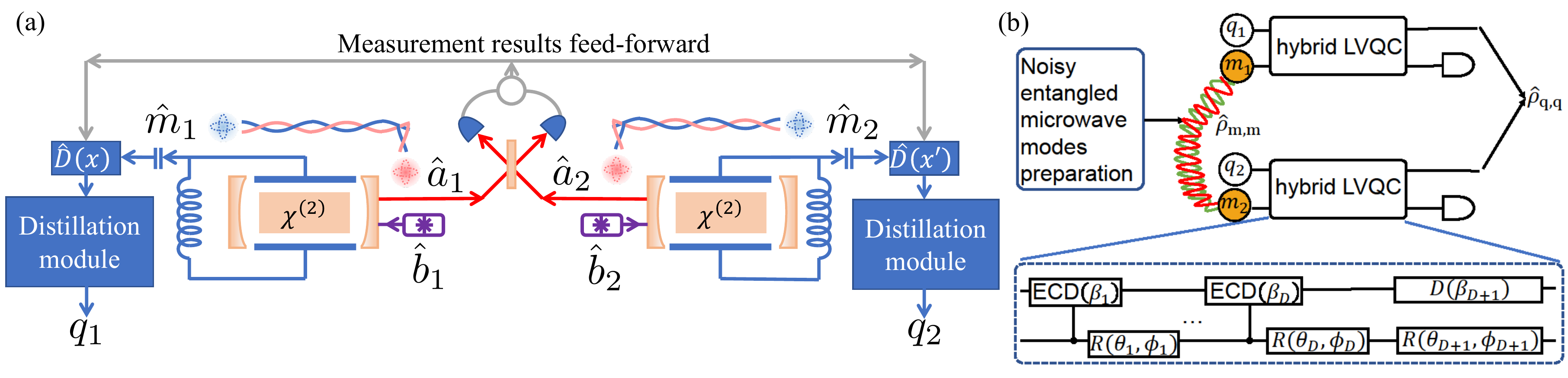}
    \caption{(a) Interconnect system between two microwave quantum computers. Two cavities enable the generation of microwave-optical entanglement. The optical modes are detected for entanglement-swap, generating microwave-microwave entanglement after displacement $\hat{D}$. Finally, the transmon qubits interact with the microwave modes to generate Bell pairs. (b)
    Schematic of hybrid local variational quantum circuits (hybrid LVQCs) to distill entanglement from noisy entangled microwave modes $m_1,m_2$ to transmon qubits $q_1,q_2$. 
    The hybrid LVQC is shown in detail in the dashed box with $D$ echoed conditional displacement (ECD) blocks (ECD gates and single qubit rotations) followed by displacement and rotation in the end. 
    }
    \label{fig:swap-scheme}
\end{figure*}

In this paper, we propose an interconnect protocol based on CV entanglement-swap to connect superconducting quantum computing circuits. We show that our approach enables order-of-magnitude rate advantage versus pure DV protocols. Furthermore, we develop two entanglement distillation protocols to convert the noisy CV entanglement to high-fidelity DV entanglement, which can directly interface with transmon qubits. The first protocol adopts a direct swap for deterministic CV-to-DV conversion and then distills the entanglement through DV operations. The second protocol is based on a hybrid local variational quantum circuit (LVQC), which uses ancillary transmon qubits to implement controlled operations on each microwave mode locally. Estimated with practical non-ideal device parameters, both protocols show huge rate advantage compared with single-photon based pure DV protocols. The LVQC protocol further shows more than ten-fold improvement for the fidelity-success-probability trade-off, compared with the direct CV-to-DV conversion protocol.

{\em System setup.---}
The overall interconnect protocol consists of two steps, a CV entanglement generation step and a CV-to-DV hybrid distillation step, as shown in Fig.~\ref{fig:swap-scheme}(a). The ultimate goal is to generate high-fidelity entanglement at a high rate between two superconducting microwave quantum computers. To begin with, entanglement is generated between the transduction microwave and optical ports ($\{\hat{a}_1,\hat{m}_1\}$ and $\{\hat{a}_2,\hat{m}_2\}$) on both sides.
The optical modes $\hat{a}_1$ and $\hat{a}_2$ then travel through optical links to a center node for entanglement swap to generate noisy CV entanglement between microwave modes $\hat{m}_1$ and $\hat{m}_2$. Next, each microwave mode is coupled into a distillation module implemented with superconducting quantum computing circuits. Then high-fidelity DV entanglement can be generated between two remote microwave modes. While our protocol is general, we consider the special case of transmon qubits for the wide use in superconducting quantum computing. 
The LVQC distillation step is based on the controlled operation on the microwave modes using transmon qubits (Fig.~\ref{fig:swap-scheme}(b)), and the final DV entanglement is also between two transmon qubits ($\{{q}_1, {q}_2\}$).

For simplicity, we assume identical configuration and performance for the two quantum computing and transduction systems.

{\em Entangling microwave modes via CV swap.---} While our protocol applies to general transduction systems, we focus on cavity electro-optic systems~\cite{fan2018superconducting,Tsang2010,Tsang2011,xu2020bidirectional} for its simplicity as it does not involve intermediate excitations. A typical cavity electro-optic system is shown in Fig.~\ref{fig:swap-scheme}(a), where 
the optical cavity with $\chi^{(2)}$ nonlinearity is placed between the capacitors of a LC microwave resonator. The electric field of the microwave mode modulates the optical resonant frequency across the capacitor via changing the refractive index of the optical cavity. Due to the mixing (rectification) between optical pump and signal in $\chi^{(2)}$ material, microwave field can be generated. The interacton Hamiltonian has the standard three-wave mixing form between two optical modes ($\hat{a}_\ell$ and $\hat{b}_\ell$) and one microwave mode ($\hat{m}_\ell$)~\cite{wu2021deterministic,Tsang2010,Tsang2011,fan2018superconducting,xu2020bidirectional},
\begin{align}
    \hat{H}_\ell = i\hbar (g\hat{a}_\ell^{\dagger}\hat{b}_\ell\hat{m}^{\dagger}_\ell-g^*\hat{a}_\ell\hat{b}_\ell^{\dagger}\hat{m}_\ell),
\end{align}
with $g$ the coupling coefficient and $\ell=1,2$ for each side.
When the optical mode $\hat{b}_\ell$ is coherently pumped, the optical mode $\hat{a}_\ell$ and the microwave mode $\hat{m}_\ell$ will be entangled in a noisy two-mode squeezed vacuum (TMSV) state $\hat{\rho}_{\rm{m,o}}$, with zero mean and a covariance matrix~\cite{wu2021deterministic}
\begin{align}
&V_{\rm{m,o}} = \frac{1}{2}
\begin{pmatrix}
u \bI_2 & v \bZ_2 \\
v \bZ_2 & w\bI_2 
\end{pmatrix},
\label{eq:CM}
\end{align}
where $\bZ_2,\bI_2$ are Pauli-Z and identity matrix, and 
\begin{subequations}
\begin{align}
&u=1+\frac{8\zeta_{\rm m} [{{C}}+ n_{\rm in}(1-\zeta_{\rm m})]}{(1-{{C}})^2},
\\
&v=\frac{4\sqrt{\zeta_{\rm o}\zeta_{\rm m} {{C}}}[1+{{C}}+2 n_{\rm in}(1-\zeta_{\rm m})]}{(1-{{C}})^2},
\\
&w=1+\frac{8{{C}}\zeta_{\rm o}\left[1+n_{\rm in}\left(1-\zeta_{\rm m}\right)\right]}{(1-{{C}})^2}.
\end{align}
\label{uvw}
\end{subequations}

Here $\zeta_{\rm o}$ and $\zeta_{\rm m}$ are the extraction efficiencies for the optical and microwave mode, respectively. The cooperativity $C\propto g^2$ describes the interaction strength~\cite{wu2021deterministic}. The optical thermal noise is neglected due to its small occupation while the microwave thermal noise has non-zero mean occupation number $n_{\rm in}$. 
We also note that any additional optical transmission loss $1-\eta$ can be absorbed into $\zeta_{\rm o}$ by replacing $\zeta_{\rm o}$ with $\eta\zeta_{\rm o}$ (see Appendix~\ref{app:optical_loss}).

To generate entanglement between remote microwave modes, we adopt an entanglement-swap approach. As shown in Fig.~\ref{fig:swap-scheme}(a), consider two pairs of entangled microwave-optical modes $\{\hat{a}_1,\hat{m}_1\}$ and $\{\hat{a}_2,\hat{m}_2\}$, each with the covariance matrix Eq.~\eqref{eq:CM}. One can interfere the optical modes $\hat{a}_1$ and $\hat{a}_2$ on a balanced beamsplitter and perform homodyne detection on the output. Conditioned on the homodyne results, displacement operations will be applied to the two microwave modes respectively.
After the optical homodyne detection and conditioned displacement operation, the microwave modes $\hat{m}_1$ and $\hat{m}_2$ form a noisy TMSV state $\hat{\rho}_{\rm{m,m}}$ with the covariance matrix
\begin{align}
V_{\rm{m,m}} & =\frac{1}{2}
\begin{pmatrix}
(u-\frac{v^2}{2w}) \bI_2 & \frac{v^2}{2w} \bZ_2 \\
\frac{v^2}{2w} \bZ_2 & (u-\frac{v^2}{2w})\bI_2
\end{pmatrix},
\label{eq:CM_scheme1_M1M2}
\end{align}
In the ideal lossless case with $\zeta_{\rm o}=\zeta_{\rm m}=1$, the state $\hat{\rho}_{\rm{m,m}}$ becomes a pure TMSV state (see Appendices~\ref{App:CV-swap} and~\ref{app:entanglement}).

We first derive the ultimate rate of entanglement generation, as all later steps are local operations and classical communications (LOCC), which does not increase the entanglement rate~\cite{horodecki2009}. To characterize the distillable entanglement, we calculate the upper bound by entanglement of formation (EoF)~\cite{bennett1996concentrating} and the lower bound by reverse coherent information (RCI)~\cite{garciapatron2009}.

\begin{figure}[t]
  \centering
    \includegraphics[width=0.48\textwidth]{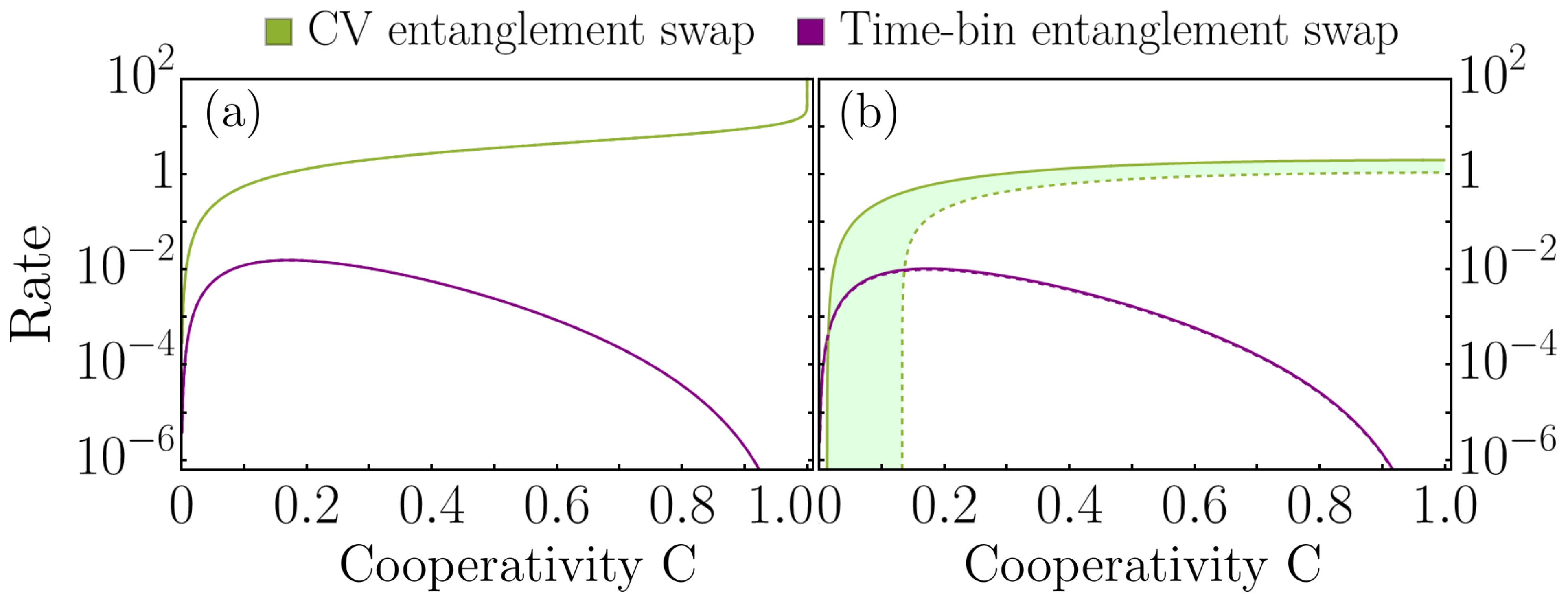}
    \caption{Rate (in ebit per round) comparison (a) $\zeta_{\rm m}=\zeta_{\rm o}=1$. (b) $\zeta_{\rm m}=0.95$, $\zeta_{\rm o}=0.9$ and $n_{\rm{in}}=0.2$. The rate of CV entanglement swap is within green shaded and the rate of time-bin entanglement swap is within purple region.
     \label{fig:line_plot}
    }
\end{figure}

To begin with, we consider the rate versus the cooperativity in Fig.~\ref{fig:line_plot}. In the ideal case $\zeta_{\rm o}=\zeta_{\rm m}=1$, EoF and RCI are equal and reduced to entanglement entropy (green line in Fig.~\ref{fig:line_plot}a). For comparison, we also evaluate the EoF and RCI rate bounds of a pure DV protocol based on the time-bin entanglement~\cite{zhong2020proposal}, where microwave-optical single-photon entanglement is generated by post-selecting on the state $\hat{\rho}_{\rm{m,o}}$ (See Appendix~\ref{App:timebin}). For all cooperativity values, the proposed CV scheme has more than two order-of-magnitude rate advantage over the pure DV protocol based on time-bin entanglement. 
In Fig.~\ref{fig:line_plot}b, we future consider the non-ideal extraction with $\zeta_{\rm m}=0.95$ and $\zeta_{\rm o}=0.9$, and non-zero microwave thermal noise $n_{\rm{in}}=0.2$ (corresponding to $\sim0.2$ Kelvin temperature at 8 GHz). 
Although rigorous advantage only happens at cooperativity above $~0.12$, we expect the lower bound to be non-tight and actual advantage should still be large in the low cooperativity region. In Appendix~\ref{app:optical_loss}, we show that such rate advantage can be identified for smaller $\zeta_{\rm o}$, and therefore is robust against optical transmission loss.
We note a recent paper~\cite{zhong2022quantum} showing similar rate advantage with a different platform and approach.

{\em Distillation protocols.---}
Next, we design distillation protocols to obtain high-fidelity DV entanglement $\hat{\rho}_{\rm{q,q}}$ between two transmon qubits from the noisy two-mode squeezing between microwave modes $\hat{\rho}_{\rm{m,m}}$.
Distillation has been explored separately for either DV~\cite{bennett1996concentrating,bennett1996purification,deutsch1996,rozpkedek2018optimizing,zhao2021practical,pan2001entanglement,kwiat2001experimental,yamamoto2003experimental,reichle2006experimental} or CV~\cite{ralph2009nondeterministic,pegg1998,takahashi2010entanglement,zhang2011,datta2012compact,vcernotik2012displacement,he2021noiseless,he2021noiseless,hu2017continuous,mardani2020continuous,campbell2013continuous,ulanov2015undoing} entanglement.
Here, we require a hybrid distillation protocol that produces high fidelity DV entanglement from noisy CV entanglement. We present two different approaches: (i) direct swap where conversion and distillation are separate; and (ii) hyrid LVQC where universal control completes both conversion and distillation simultaneously.
With symmetric configuration between the two ends of the interconnect system, we will omit subscripts in the following.

{\em Direct swap.---} We begin with a simple direct swap from CV to noisy entangled qubit pairs. Inspired by Ref.~\cite{agusti2022long}, we consider the interaction between the microwave mode and a transmon qubit with Hamiltonian
\be 
\hat{H}_{\rm swap}=\hat{m}\otimes \ketbra{1}{0}_{q}+h.c.,
\label{eq:direct_swap_H}
\ee
We can control the interaction time $t$ such that the unitary operator $\hat{U}=\exp(-it\hat{H}_{\rm swap})$ gives the maximum EoF between the qubits from the two sides after disregarding the corresponding microwave modes (see Appendix~\ref{appendix:direct_swap}). 
This will produce a noisy entangled qubit pair deterministically. One can then follow up with further DV distillation protocols.

{\em Variational hybrid distillation.---}
To explore the ultimate performance, we further use a variational approach to design a hybrid distillation protocol, as shown in Fig.~\ref{fig:swap-scheme}(b). Two qubits are initialized in $\ket{0}_q$, and put into interaction with the two microwave modes respectively. The hybrid LVQC includes a series of single-qubit rotation and echoed conditional displacement (ECD), followed by another displacement at the end~\cite{eickbusch2021fast}.
Each single qubit rotation is characterized by two angles $\phi$ and $\theta$,
\be 
\hat{R}(\theta,\phi) = \exp[-i(\theta/2)(\hat{\sigma}_x\cos\phi + \hat{\sigma}_y \sin\phi)]
\ee
where $\hat{\sigma}_x$ and $\hat{\sigma}_y$ are the Pauli operators. Each ECD gate acts on the microwave mode $\hat{m}$ and the qubit $\hat{q}$ as
\be 
\hat{\rm ECD}(\beta) = \hat{D}\left(\beta/2\right)\otimes \ket{1}\bra{0}_{q} + h.c., 
\ee 
where $\hat{D}(\alpha) \equiv \exp(\alpha \hat{m}^\dagger-\alpha^* \hat{m})$ is the displacement operator. Compared with previous proposals for universal control~\cite{heeres2015,krastanov2015universal}, the considered ECD gate approach has advantage in both control speed and fidelity~\cite{eickbusch2021fast}.

To determine the success of entanglement distillation, we perform measurements characterized by positive operator-valued measure (POVM) $\{\hat\Pi_{\rm s}, \hat{\bI}-\hat\Pi_{\rm s}\}$. The success probability is therefore
\begin{equation}
    P_{\rm success} = \Tr\left\{\hat\Pi_{\rm s}\left(\hat U_D\otimes \hat U_D\right)\hat{\rho}_{\rm qm,qm}\left( \hat U_D^\dagger\otimes \hat U_D^\dagger\right)\right\},
    \label{eq:psucc}
\end{equation}
where $\hat{\rho}_{\rm qm,qm}$ is the composite initial quantum state including both microwave modes and qubits on the two sides and $\hat{U}_D$ is the unitary of the LVQC with $D$ layers of ECD and single-qubit rotations. 
We choose photon counting on the microwave modes and post-select on photon number lower than certain threshold $n_{\rm th}$, so $\Pi_{\rm s}=\sum_{i,j=0}^{n_{\rm th}-1} \state{i}_{m_1}\otimes\state{j}_{m_2}$, where $\ket{k}$ is the number state. In this paper, we choose $n_{\rm th}=5$, which is half of the photon number cut off in simulation.
The state of transmon qubits conditioned on successful distillation is given by
\begin{equation}
    \hat{\rho}_{\rm q,q} \propto\Tr_{\rm m,m}\left\{\hat\Pi_{\rm s}\left(\hat{U}_D\otimes \hat{U}_D\right)\hat{\rho}_{\rm qm,qm}\left(\hat{U}_D^\dagger\otimes \hat{U}_D^\dagger\right)\hat\Pi_{\rm s} \right\}.
    \label{eq:rho_qq}
\end{equation}

We train the hybrid LVQC towards distilling a perfect Bell pair $\ket{\Psi^+}\equiv (\ket{0}_q\ket{0}_q+\ket{1}_q\ket{1}_q)/\sqrt{2}$. The cost function is defined with the success probability $P_{\rm success}$ and fidelity $F\left(\hat{\rho}_{\rm qq}\right) = \bra{\Psi^+}\hat{\rho}_{\rm q,q}\ket{\Psi^+}$, as
\begin{equation}
    \calC\left(\{\beta\},\{\theta\},\{\phi\}\right) = (1-P_{\rm success}) + \lambda\times {\rm Softplus}(F_c - F),
    \label{eq:cost}
\end{equation}
where the penalty coefficient $\lambda$ and critical fidelity $F_c$ are hyperparameters to tune the tradeoff between success probability and fidelity. The softplus function ${\rm Softplus}(x)\equiv \log\left(1+e^{\gamma x}\right)/\gamma$ with $\gamma=20$ is introduced to enable a smooth penalty on the $F\le F_c$ events. 

\begin{figure}[t]
    \centering
    \includegraphics[width=0.45\textwidth]{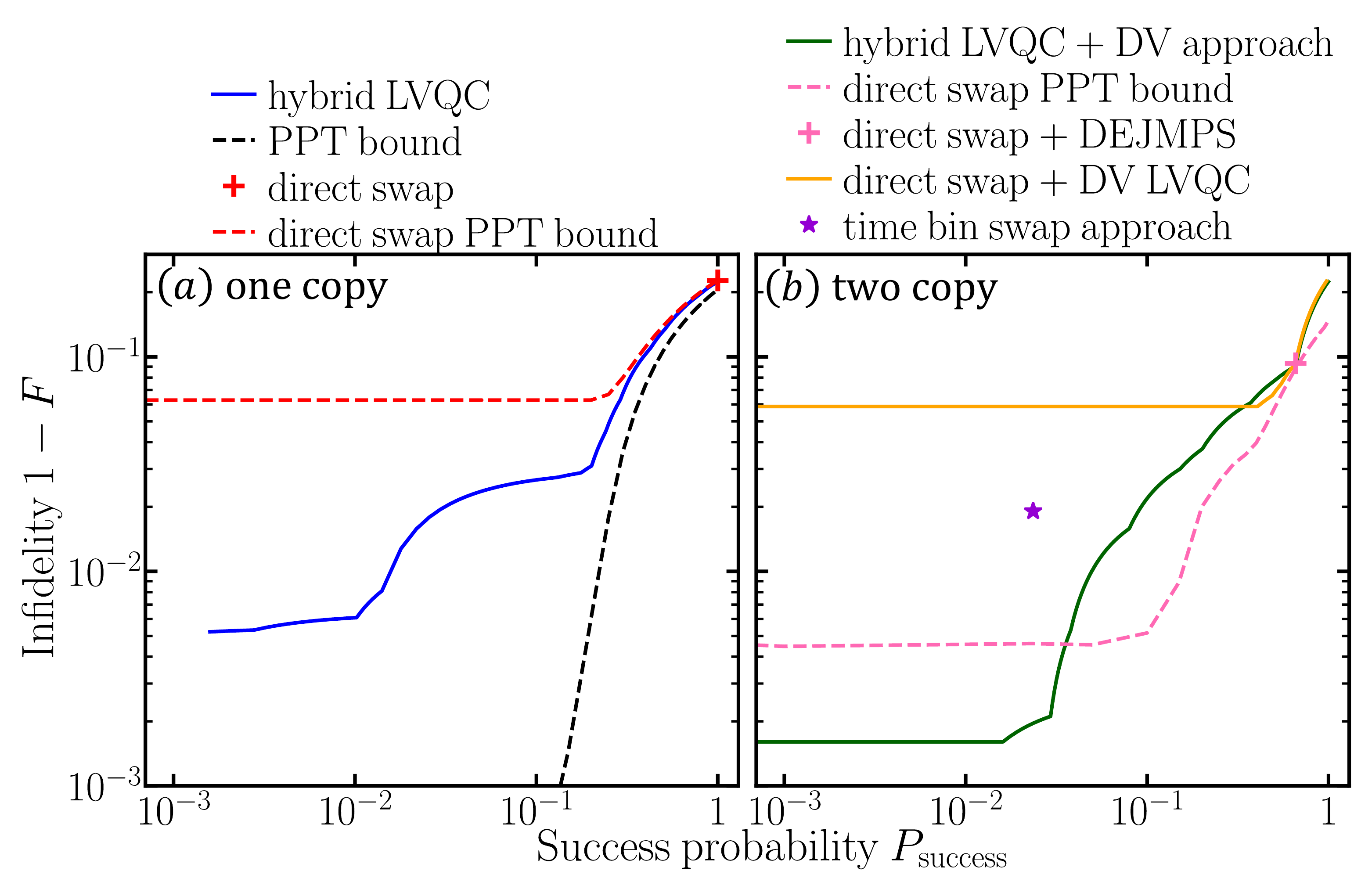}
    \caption{Infidelity of transmon qubits versus success probability on (a) one-copy and (b) two-copy entangling microwave modes with $C=0.1, n_{\rm in}=0.2, \zeta_{\rm m}=0.992, \zeta_{\rm o}=0.99$. 
    The PPT bound (black) in (a) goes below $10^{-4}$ and saturates; red and pink dashed curves in (a)(b) correspond to PPT bound of one-copy and two-copy two-qubit state by direct swap. There are $D=10$ ECD blocks in hybrid LVQCs and $L=6$ layers in DV LVQCs. 
    }
    \label{fig:distillation_performance}
\end{figure}

Both the direct swap and hybrid LVQC can be operated on a single copy of $\hat{\rho}_{\rm{m,m}}$ to generate a single copy ($M=1$) of entangled qubit pair. To further improve their performance, we also consider a two-copy case ($M=2$) where two copies of noisy CV entanglement $\hat{\rho}_{\rm{m,m}}$ are first converted to two pairs of noisy entangled qubits, and then further DV distillation is performed on the two noisy qubit pairs to produce the final entangled state. In the last step, we consider both the traditional DEJMPS protocol~\cite{deutsch1996} and a DV LVQC protocol. The DV LVQC protocol utilizes a standard universal gate set of single-qubit rotations and CNOTs, with Pauli-Z measurement providing the post-selection and the same cost function as Eq.~\eqref{eq:cost}, as we detail in Appendix~\ref{appendix:qubit_distill}.

{\em Performance comparison.---}
We begin our performance comparison with the trade-off between infidelity $1-F$ and the success probability $P_{\rm success}$. As shown in Fig.~\ref{fig:distillation_performance}(a), the simple direct swap produces $1-F\sim 0.23$ deterministically (red cross). The performance of any protocol following the direct swap will be bounded by the positive-partial-transpose (PPT) bound (red dashed), which can be numerically evaluated by a semidefinite program~\cite{rozpkedek2018optimizing}. In contrast, the hybrid LVQC approach directly achieves an one-order-of-magnitude advantage in the infidelity (blue solid). Indeed, when applying the PPT bound on $\hat{\rho}_{\rm m,m}$, the (generally loose) lower bound  of infidelity for LVQC protocols (black dashed) also decreases substantially, compared with the PPT bound from the direct swap (red dashed).

To further lower the infidelity, we consider the two-copy case, where DV processing is further performed on the output of the one-copy case. We first consider the traditional DEJMPS following the direct swap protocol as the reference (pink cross in Fig.~\ref{fig:distillation_performance}(b)). When the traditional DEJMPS is replaced by DV LVQC (orange line), lower infidelity can be achieved with lower success probability. When we use DV LVQC to further distill the two qubit pairs from the hybrid LVQC, a two-order-of-magnitude lower infidelity (green) can be achieved.
Note that even compared with the (generally loose) two-copy PPT lower bound of the direct swap approach (pink dashed), advantages can still be identified in the low infidelity region. As a benchmark, the time-bin based entanglement swap (see Appendix~\ref{App:timebin}) (purple star) is one-order-of-magnitude worse in terms of the infidelity, when compared with the hybrid LVQC (green) at the same success probability.

\begin{figure}[t]
    \centering    \includegraphics[width=0.48\textwidth]{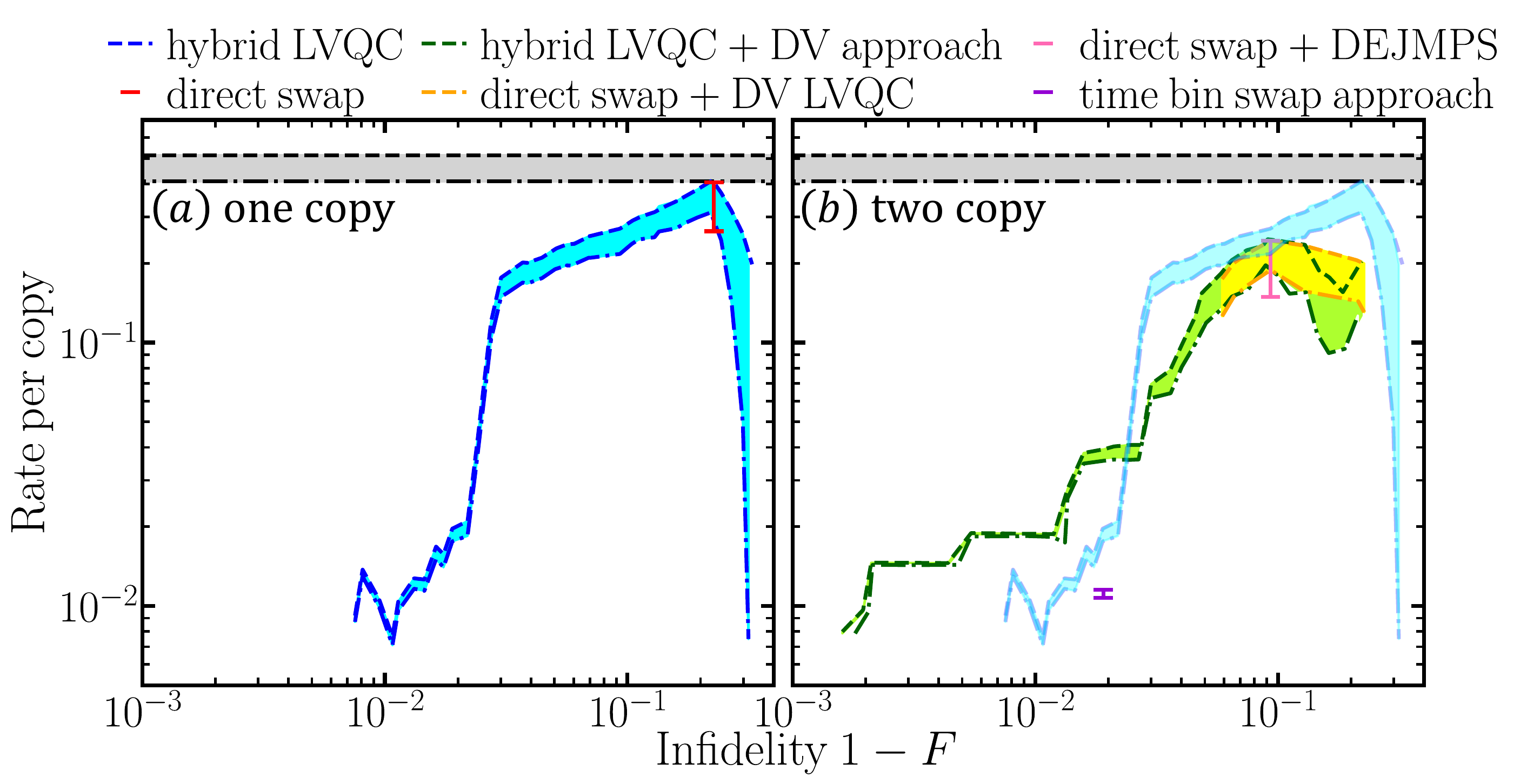}
    \caption{Entanglement distillation rate per copy on (a) one-copy and (b) two-copy noisy entangled microwave modes versus infidelity $1-F$, at identical parameters to Fig.~\ref{fig:distillation_performance}. Dot-dashed and dashed curves correspondingly represent RCI lower bounds and EoF upper bounds. Shaded areas and line segments show the range of rate. The shallow blue curves and areas in (b) are same as (a) for comparison. 
    \label{fig:distillation_rate}
    }
\end{figure}

Although we only considered two copies, the advantage of the hyrid LVQC scheme over the direct swap scheme can be generalized to more copies. As the hyrid LVQC protocol starts with better entanglement generation performance (Fig.~\ref{fig:distillation_performance}(a)), it will also perform better when further distillation steps are implemented.

Finally, we compare the distillation rate per copy for a $M$-copy distillation protocol,
\be
R_E(\hat{\rho}_{\rm q,q}) = P_{\rm success}\times E\left(\hat{\rho}_{\rm q,q}\right)/M \le E\left(\hat{\rho}_{\rm m,m}\right),
\ee
where $E\left(\cdot\right)$ is the distillable entanglement. As we discussed, we utilize the lower and upper bounds of distillable entanglement, RCI and EoF for evaluation. In Fig.~\ref{fig:distillation_rate}, we see that the rate in general decays as the required infidelity decreases; this is because we are considering a fixed number of copies, which is far from an asymptotic limit. In the one-copy case, the hybrid LVQC approach (blue) is able to achieve close to optimal rate while improving the fidelity up to $0.97$, while further improving fidelity drastically decreases the rate; at the same time, direct swap (red) only achieves a single point with a large infidelity and a similar rate, compared with a versatile rate-infidelity trade-off enabled by LVQC. In the two-copy case, hybrid LVQC with additional DV post-processing is able to improve the infidelity to $\sim10^{-3}$ while the rate is kept $\sim10^{-2}$ (green). At higher infidelity, hybrid LVQC is also as good as the other approaches (orange) with direct swap and additional DV post-processing (including a DV LVQC circuit). Most notably, at the same level of infidelity, the rate of the hybrid LVQC approach (green) is a factor of $3.4$ higher than the time-bin entanglement swap approach (purple dot), as shown in Fig.~\ref{fig:distillation_rate}(b).

In Fig.~\ref{fig:distillation_rate}(b), we also compare the two-copy results with the one-copy hybrid LVQC results (shallow blue). It is interesting to note that the one-copy protocol has higher rates for the same infidelity when $1-F\gtrsim 0.024$. This is mainly due to the fact that in a two-copy protocol we need to post-select on two one-copy success events, which substantially reduces the overall success probability. However, if one wants to reach substantially lower infidelity, then two-copy protocols give a much better rate.

{\em Discussions.---}We propose an interconnect system based on CV entanglement swap and hybrid variational entanglement distillation to entangle microwave superconducting quantum computers. Our approach provides a huge rate and fidelity advantage compared with time-bin approach. In particular, the hybrid distillation protocol provides an infidelity-success-probability trade-off, with order-of-magnitude advantage over the direct swap approach. In the multi-copy scenario, it is an open problem how to optimize the circuit design to get closer to the PPT lower bound.
Although we have considered cavity electro-optics as the transduction system, our protocol also applies to other transduction systems and the analyses can be done in the same way based on a modified interaction Hamiltonian Eq.~\eqref{eq:CM}.

\begin{acknowledgements}
QZ and LF acknowledge discussion with Filip Rozp\c{e}dek, Chi Xiong, Edward H Chen, Jason S Orcutt, John A Smolin, Vikesh Siddhu and Abram L Falk.
BZ, JW and QZ acknowledge support from NSF CAREER Award CCF-2142882, Defense Advanced Research Projects Agency (DARPA) under Young Faculty Award (YFA) Grant No. N660012014029 and National Science
Foundation (NSF) Engineering Research Center for
Quantum Networks Grant No. 1941583. FL acknowledges support from U.S. Department of Energy UT-Battelle/Oak Ridge National Laboratory (4000178321), Office of Naval Research (N00014-19-1-2190) and National Science Foundation (CCF-1907918).
\end{acknowledgements}

\appendix

\section{Continuous-variable entanglement swap}
\label{App:CV-swap}
In this paper, we adopt the same convention of quadratures and covariance matrix as Ref.~\cite{wu2021deterministic}. For a review of these definitions, please refer to appendices of Ref.~\cite{wu2021deterministic}. Below we give a simple overview. The Wigner function of a Gaussian state $\hat{\rho}$ is 
\begin{equation}
    W(\bx) = \frac{1}{(2\pi)^n \vert\boldsymbol{V}\vert^{\frac{1}{2}}} \exp[-\frac{1}{2}(\bx - \Bar{\bx})^{\rm T}\boldsymbol{V}^{-1}(\bx-\Bar{\bx})],
    \label{eq:Wigner-Gaussian-state}
\end{equation}
where $\Bar{\bx}$ and $\boldsymbol{V}$ are the mean and covariance matrix of state $\hat{\rho}$:
\begin{align}
    & \Bar{\bx} \equiv \text{Tr}[\hat{\rho} \hat{\bx}],\\
    & V_{ij} \equiv
    \frac{1}{2}\text{Tr}[\hat{\rho} \{\hat{x}_i-\bar{x}_i,\hat{x}_j-\bar{x}_j\}].
\end{align}
Here $\hat{\bx}\equiv (\hat{q}_1,\hat{p}_1,\hat{q}_2,\hat{p}_2,\cdots)$ is the vector of quadrature operators.

A Gaussian state is mapped to another Gaussian state under a Gaussian unitary $\hat{U}_{\boldsymbol{S},\boldsymbol{d}}$ (described by a symplectic matrix $\boldsymbol{S}$ and a displacement vector $\boldsymbol{d}$), with the mean and covariance matrix transformed as
\begin{equation}
    \bar{\bx}\rightarrow\boldsymbol{S}\bar{\bx}+\boldsymbol{d},\; \boldsymbol{V}\rightarrow\boldsymbol{S}\boldsymbol{V}\boldsymbol{S}^{\rm T}.
    \label{eq:Symplectic_transform_Gaussian}
\end{equation}

\subsection{Entanglement swap}

\begin{figure}[t]
    \centering
    \includegraphics[width=0.45\textwidth]{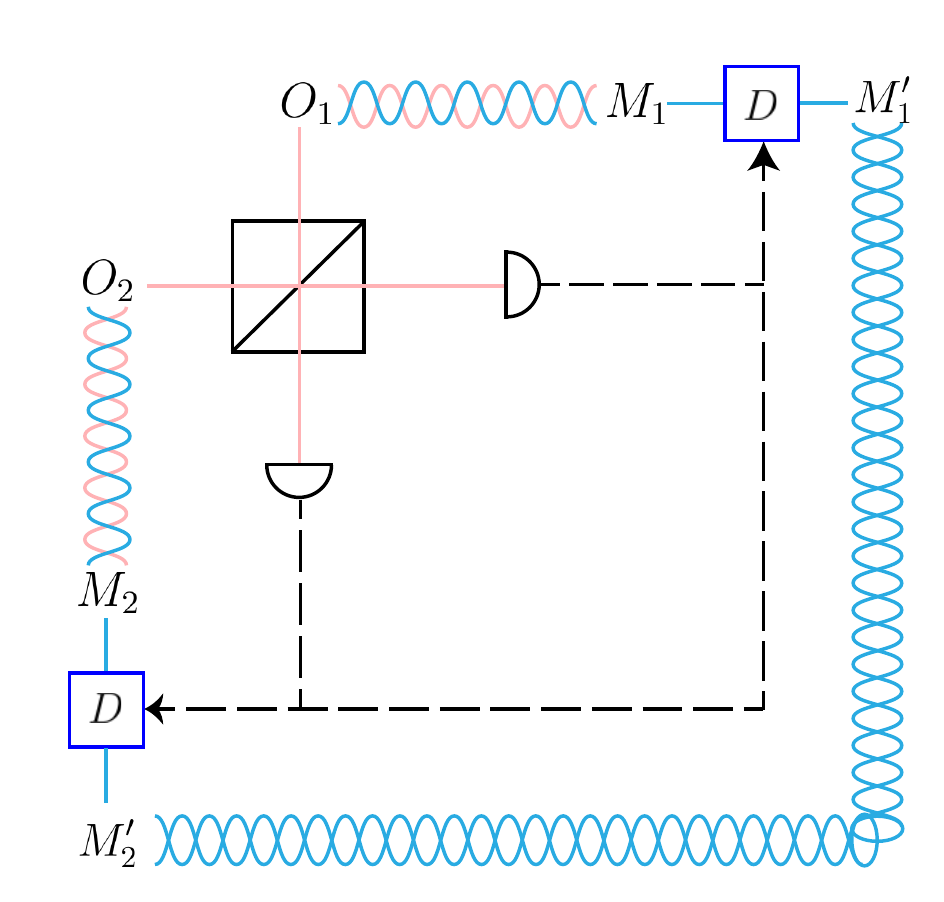}
    \caption{Continuous-variable  entanglement swap scheme. The cavities generate two entangled mode pairs $O_1M_1$ and $O_2M_2$. Then two optical modes $O_1$ and $O_2$ are mixed by a balanced beam splitter. We perform homodyne measurements at the output of beam splitter and then displace the microwave modes $M_1$ and $M_2$, producing two entangled microwave modes $M_1'$ and $M_2'$.}
    \label{fig:CV-swap-scheme}
\end{figure}
The continuous-variable (CV) entanglement scheme is shown in Fig.~\ref{fig:CV-swap-scheme}. Each of the entangled mode pairs $O_1M_1$ (with annihilation operators $\{\hat{a}_1,\hat{m}_1\}$) and $O_2M_2$ (with annihilation operators $\{\hat{a}_2,\hat{m}_2\}$) from the cavities are in a noisy TMSV state $\hat{\rho}_{\rm{m,o}}$, with zero mean and the covariance matrix $V_{\rm{m,o}}$ given by Eq.~\eqref{eq:CM} in the main text. 

The composite four-mode system $M_1M_2O_1O_2$ in the entanglement swap starts in a product of two identical Gaussian states $\hat{\rho}_{\rm{m,o}}$ with an overall covariance matrix 
\begin{align}
    V_{M_1M_2O_1O_2} & = \frac{1}{2}
    \begin{pmatrix}
    u\bI_2 & \mathbf{0} & v\bZ_2 & \mathbf{0}\\
    \mathbf{0} & u\bI_2 & \mathbf{0} &v\bZ_2\\ 
    v\bZ_2 & \mathbf{0} & w\bI_2 & \mathbf{0}\\
    \mathbf{0} & v\bZ_2 & \mathbf{0} & w\bI_2
    \end{pmatrix}.
    \label{eq:CM_composite_system}
\end{align}
First, one interferes the optical modes $O_1$ (with annihilation operator $\hat{a}_1$) and $O_2$ (with annihilation operator $\hat{a}_2$) with a balanced beam splitter, described by the symplectic transform 
\begin{equation}
    \mathbf{S}_{\rm{BS}} = \bI_4 \otimes \frac{1}{\sqrt{2}}
    \begin{pmatrix}
    \bI_2 & \bI_2 \\
    -\bI_2 & \bI_2 
    \end{pmatrix}.
\end{equation}
The covariance matrix after the beam splitter
\begin{align}
 V_{M_1 M_2 O_1 O_2}^\prime & =             \mathbf{S}_{\rm{BS}} V_{M_1M_2O_1O_2}
    \mathbf{S}_{\rm{BS}}^{\rm T} \equiv
    \begin{pmatrix}
    \mathbf{C}_1 & \mathbf{C}_3\\
    \mathbf{C}^{\rm T}_3 & \mathbf{C}_2
    \end{pmatrix},
\end{align}
where we have denoted it in a block form. The position quadrature $\hat{q}_-$ and the momentum quadrature $\hat{p}_+$ of the outputs of beam splitter  are measured by homodyne. Let $\tilde{\bx}=\sqrt{2}(q_{-},-p_{+})^{\rm T}$ be the rescaled result of homodyne measurements, and $\Pi_4=\text{Diag}(0,1,1,0)$ be the projection to integrate out the other variables in the Wigner function. The state of $M_1 M_2$ after the homodyne measurements is described by its covariance matrix
\begin{align}
V_{M_1M_2} & = \mathbf{C}_1-\mathbf{C}_3(\Pi_4\mathbf{C}_2\Pi_4)^{-1}\mathbf{C}_3^{\rm T}, \nonumber \\
& =\frac{1}{2}
\begin{pmatrix}
(u-\frac{v^2}{2w}) \bI_2 & \frac{v^2}{2w} \bZ_2 \\
\frac{v^2}{2w} \bZ_2 & (u-\frac{v^2}{2w})\bI_2
\end{pmatrix},
\label{eq:CM_CV_M1M2}
\end{align}
and mean 
\begin{align}
    \bar{\bx}_{12}
    =(-\frac{v}{2w}\bZ_2 \tilde{\bx},\frac{v}{2w}\tilde{\bx}).
\end{align}
When displacements $\hat{D}(\frac{v}{2w} \bZ_2 \tilde{\bx})$ and $\hat{D}(-\frac{v}{2w} \tilde{\bx})$ are applied at modes $M_1$ and $M_2$, respectively, the final output state of system $M_1'M_2'$ (with annihilation operators $\hat{m}_1,\hat{m}_2$) is a zero-mean Gaussian state with the same covariance matrix as $V_{M_1M_2}$ in Eq.~\eqref{eq:CM_CV_M1M2}. (For the definition of the Weyl displacement operator please refer to \cite{wu2021deterministic}.)

\begin{figure}[t]
    \centering
    \includegraphics[]{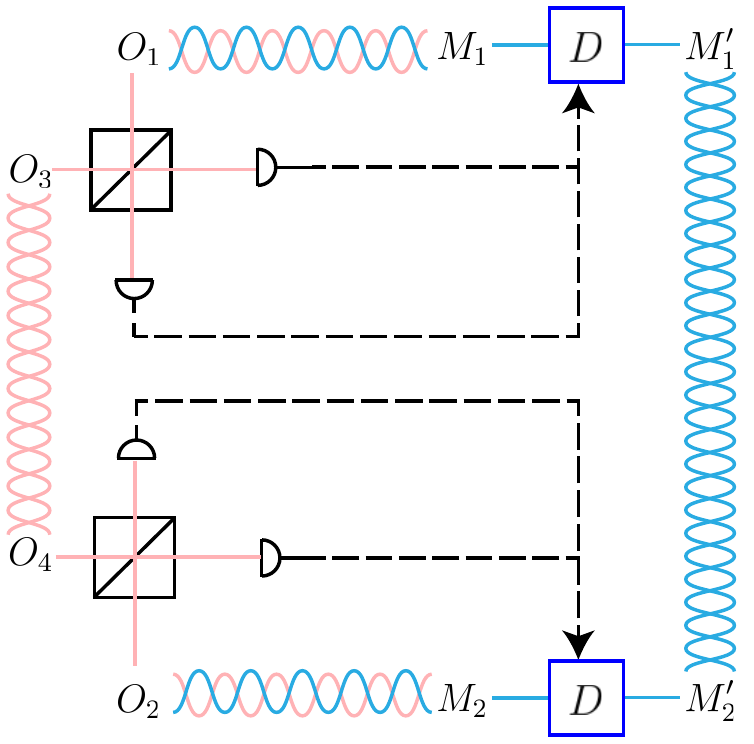}
    \caption{CV teleportation scheme. We start with a TMSV state between optical mode $O_3$ and $O_4$. Then they will be teleported by use of two M-O pairs. After teleportation, the microwave modes $M_1'$ and $M_2'$ are entangled.}
    \label{fig:CV-teleportate-scheme2}
\end{figure}

\subsection{Teleportation approach}
The other CV-teleportation approach is shown in Fig.~\ref{fig:CV-teleportate-scheme2}. 
In this approach, one starts with a strong TMSV pair $O_3O_4$ in the optical domain, with covariance matrix
\begin{equation}
    V_{O_3O_4} = \frac{1}{2} 
    \begin{pmatrix}
    \cosh(2r) \bI_2 & \sinh(2r) \bZ_2\\
    \sinh(2r) \bZ_2 & \cosh(2r) \bI_2
    \end{pmatrix}.
\end{equation}
The covariance matrix of the composite system  $M_1 M_2 O_1 O_3 O_2 O_4$ is 
\begin{equation}
    V= \frac{1}{2}\begin{pmatrix}
    u \bI_2 & \mathbf{0} & v\bZ_2 & \mathbf{0} & \mathbf{0} & \mathbf{0}\\
    \mathbf{0} & u\bI_2 & \mathbf{0} & \mathbf{0}& v\bZ_2 & \mathbf{0} \\
    v\bZ_2 & \mathbf{0}& w\bI_2 & \mathbf{0} & \mathbf{0} & \mathbf{0}\\
    \mathbf{0} & \mathbf{0} & \mathbf{0} &\cosh{(2r)}\bI_2 & \mathbf{0} &\sinh{(2r)}\bZ_2\\
    \mathbf{0} & v\bZ_2 & \mathbf{0}& \mathbf{0}&w\bI_2& \mathbf{0}\\
    \mathbf{0} & \mathbf{0}& \mathbf{0}&\sinh{(2r)}\bZ_2 & \mathbf{0} &\cosh{(2r)} \bI_2
    \end{pmatrix}.
\end{equation}
Two balanced beam splitters are applied to $O_1O_3$ and $O_2O_4$ separately:
\begin{align*}
    \mathbf{S}_{\rm{BS}1} &= \bI_2 \otimes \frac{1}{\sqrt{2}}
    \begin{pmatrix}
    \bI_2 & \bI_2 \\
    -\bI_2 & \bI_2 
    \end{pmatrix} \otimes \bI_2,\\
    \mathbf{S}_{\rm{BS}2} &= \bI_4 \otimes \frac{1}{\sqrt{2}}
    \begin{pmatrix}
    \bI_2 & \bI_2 \\
    -\bI_2 & \bI_2 
    \end{pmatrix},
\end{align*}
leading to the output
\begin{align}
V^\prime &= \mathbf{S}_{\rm{BS}2}\mathbf{S}_{\rm{BS}1}V\mathbf{S}_{\rm{BS}1}^{\rm{T}}\mathbf{S}_{\rm{BS}2}^{\rm{T}},
\equiv  
    \begin{pmatrix}
    \mathbf{D}_1 & \mathbf{D}_3\\
    \mathbf{D}^{\rm T}_3 & \mathbf{D}_2
    \end{pmatrix}.
\end{align}
Let $\Pi_8\equiv \text{Diag}(0,1,1,0,0,1,1,0)$ be the projection, $\tilde{\bx}^{(1)}\equiv \sqrt{2}(q_{-}^{(3)},-p_{+}^{(1)})^{\rm T}$ be measurement results on $O_1O_3$ port and $\tilde{\bx}^{(2)}\equiv \sqrt{2}(q_{-}^{(4)},-p_{+}^{(2)})^{\rm T}$ be measurement results on $O_2O_4$ port.
Then the covariance matrix and mean of the output microwave modes $M_1 M_2$ are 
\begin{widetext}
\begin{align}
V_{12} & = \mathbf{D}_1-\mathbf{D}_3(\Pi_8\mathbf{D}_2\Pi_8)^{-1}\mathbf{D}_3^{\rm T} =\frac{1}{2}
\begin{pmatrix}
(u-\frac{v^2[w+\cosh{(2r)}]}{1+w^2+2w\cosh{(2r)}}) \bI_2 & \frac{v^2\sinh{(2r)}}{1+w^2+2w\cosh{(2r)}} \bZ_2 \\
\frac{v^2\sinh{(2r)}}{1+w^2+2w\cosh{(2r)}} \bZ_2 & (u-\frac{v^2[w+\cosh{(2r)}]}{1+w^2+2w\cosh{(2r)}})\bI_2
\end{pmatrix}, 
\label{eq:CM_scheme2_M1M2}
\\
\bar{\bx}_{12}&
=(-\bZ_2\tilde{\bx},\tilde{\bx})
, \mbox{ with }
\tilde{\bx} =\frac{v\sinh{(2r)}}{1+w^2+2w\cosh{(2r)}}\tilde{\bx}^{(1)}-\frac{v[w+\cosh{(2r)}]}{1+w^2+2w\cosh{(2r)}}\bZ_2\tilde{\bx}^{(2)}.
\label{eq:mean_scheme2_M1M2}
\end{align}
\end{widetext}
We can then perform displacements to cancel the mean and generate the final entangled microwave system $M_1'M_2'$, with zero-mean and covariance matrix given by Eq.~\eqref{eq:CM_scheme2_M1M2}.
The output of the second scheme reduces to that of the first scheme when $O_3O_4$ are in the infinite squeezing limit ($r\rightarrow +\infty$). 

\section{Entanglement measures}
\label{app:entanglement}

The evaluation of distillable entanglement in general requires a regularization of an infinite number of copies; therefore, as explained in the main text, we consider the lower and upper bounds of it, RCI and EoF~\cite{vedral1998entanglement,brandao2008entanglement}. For a bipartite quantum system with $A$ and $B$ in a state $\hat{\rho}_{AB}$, if classical communication is allowed, then the rate of entanglement generation is lower bounded by RCI~\cite{garciapatron2009}
\begin{equation}
    I_R(\hat{\rho}_{AB}) = \max\{S\left(\hat{\rho}_B\right), S\left(\hat{\rho}_A\right)\} - S\left(\hat{\rho}_{AB}\right),
    \label{eq:RCInfo}
\end{equation}
where $S(\hat{\rho})=-\Tr{\hat{\rho}\log_2\hat{\rho}}$ is the von Neumman entanglement entropy of state $\hat{\rho}$ and $\hat{\rho}_A=\Tr_{B}\hat{\rho}_{AB}$ is the reduced density matrix of subsystem $A$ and  similarly for $\hat{\rho}_B$.

For a symmetric two-mode Gaussian state characterized by the covariance matrix
\begin{equation}
    V = \frac{1}{2}\begin{pmatrix}
    a\bI_2 & c\bZ_2\\
    c\bZ_2 & b\bI_2
    \end{pmatrix},
    \label{eq:CM_modes}
\end{equation}
in the standard form, its symplectic eigenvalues are simply given as $\nu_{\pm} = [\sqrt{y} \pm (b-a)]/2$ with $y \equiv (a+b)^2-4c^2$~\cite{Weedbrook_2012}.
The entanglement entropy is
$
S(\hat{\rho}) = g(\nu_+) + g(\nu_-),
$
where 
\be 
g(x)\equiv \frac{x+1}{2}\log_2\left(\frac{x+1}{2}\right) - \frac{x-1}{2}\log_2\left(\frac{x-1}{2}\right).
\ee 

For the microwave modes in Eq.~\eqref{eq:CM_scheme1_M1M2} of the main text, its symplectic eigenvalues are equal
\begin{equation}
    \nu_{\pm} = \sqrt{u\left(u-\frac{v^2}{w}\right)},
\end{equation}
and the entropy of the entire system is 
\begin{equation}
    S(\hat{\rho}_{\rm mm}) = 2g\left(\sqrt{u\left(u-\frac{v^2}{w}\right)}\right).
\end{equation}
The RCI is thus
\begin{equation}
    I_R(\hat{\rho}_{\rm mm}) = g\left(u-\frac{v^2}{2w}\right) - 2g\left(\sqrt{u\left(u-\frac{v^2}{w}\right)}\right).
\end{equation}

On the other hand, EoF quantifies the resources that are required to create a quantum state in terms of the number of Bell pairs (or ``ebit''), which is formally defined as~\cite{bennett1996concentrating}
\begin{equation}
    E_f\left(\hat{\rho}\right) = \min_i \sum_i p_i S(\ket{\psi_i}),
    \label{eq:eof}
\end{equation}
where the minimum is taken over all possible pure state decomposition of state $\hat{\rho}$ as $\hat{\rho} = \sum_{i}p_i \ket{\psi_i}\bra{\psi_i}$.
Specifically, for a pure state, both RCI and EoF reduce to entanglement entropy.

EoF is in general subadditive, 
\be 
E_f\left(\hat{\rho}_1\otimes \hat{\rho}_2\right) \le E_f\left(\hat{\rho}_1\right)+E_f\left( \hat{\rho}_2\right).
\ee 

EoF is known to be analytically solvable for the symmetric two-mode Gaussian state $\hat{\rho}$ with the covariance matrix $V$ in Eq.~\eqref{eq:CM_modes} when $a=b$~\cite{tserkis2020maximum} as
\begin{equation}
    E_f(\hat{\rho}) = \begin{cases}
    h(\sqrt{\nu_1\nu_2}) & \textit{if $\nu_1\nu_2 < 1$,}\\
    0 & \textit{otherwise,}
    \end{cases}
\end{equation}
where $\nu_1, \nu_2$ are the first two eigenvalues of $V$ in increasing order and $h(x)$ is defined as
\begin{equation}
    h(x) \equiv \frac{(1+x)^2}{4x}\log_2\left(\frac{(1+x)^2}{4x}\right) - \frac{(1-x)^2}{4x}\log_2\left(\frac{(1-x)^2}{4x}\right).
\end{equation}
For the noisy entangled microwave modes (see Eq.~\eqref{eq:CM_scheme1_M1M2} of the main text), we have 
$
\nu_1 = \nu_2 = u-\frac{v^2}{w}
$
and EoF is
\begin{equation}
    E_f(\hat{\rho}_{\rm m,m}) = h\left(u-\frac{v^2}{w}\right).
\end{equation}

EoF is analytically solvable for arbitrary two-qubit states~\cite{wootters1998entanglement} and we explain it in the following. We first introduce the spin-flipped state of $\hat{\rho}$ as
\begin{equation}
    \hat{\tilde{\rho}} = (\hat\sigma_y\otimes \hat\sigma_y)\hat{\rho}^*(\hat\sigma_y\otimes \hat\sigma_y),
\end{equation}
where $\hat{\rho}^*$ is the complex conjugate of $\hat{\rho}$. Define the function $\calE(x)$ as
\begin{equation}
    \calE(x) = H\left(\frac{1+\sqrt{1-x^2}}{2}\right),
\end{equation}
where $H(x)\equiv -x\log_2(x) -(1-x)\log_2(1-x)$ is the binary entropy. Note that the $\calE$ function monotonically increases with $x$ from $0$ to $1$. The EoF for an arbitrary two-qubit state $\hat{\rho}$ is thus
\begin{equation}
    E_f(\hat{\rho}) = \calE\left[\delta\left(\hat{\rho}\right)\right], 
\end{equation}
where $\delta(\hat{\rho}) \equiv \max\{0, v_1-v_2-v_3-v_4\}$ and $\{v_i\}_{i=1}^4$ are the square root of eigenvalues of $\hat{\rho}\hat{\tilde{\rho}}$ in decreasing order.

In the main text, we evaluate the entanglement measures for the noisy two-mode squeezed vacuum in Eq.~\eqref{eq:CM_scheme1_M1M2} of the main text.
When $\zeta_{\rm o}=\zeta_{\rm m}=1$, $\hat{\rho}_{\rm{m,m}}$ is in a pure TMSV state with mean photon number
$ 
N_{\rm Ideal}={16C^2}/[(1-C)^2 (1+6C+C^2)].
$
In this pure state case, the EoF and RCI are both equal to $g\left(2N_{\rm Ideal}+1\right)$.

\section{Time-bin entanglement swap}
\label{App:timebin}
\begin{figure}[t]
    \centering
    \includegraphics[width=0.45\textwidth]{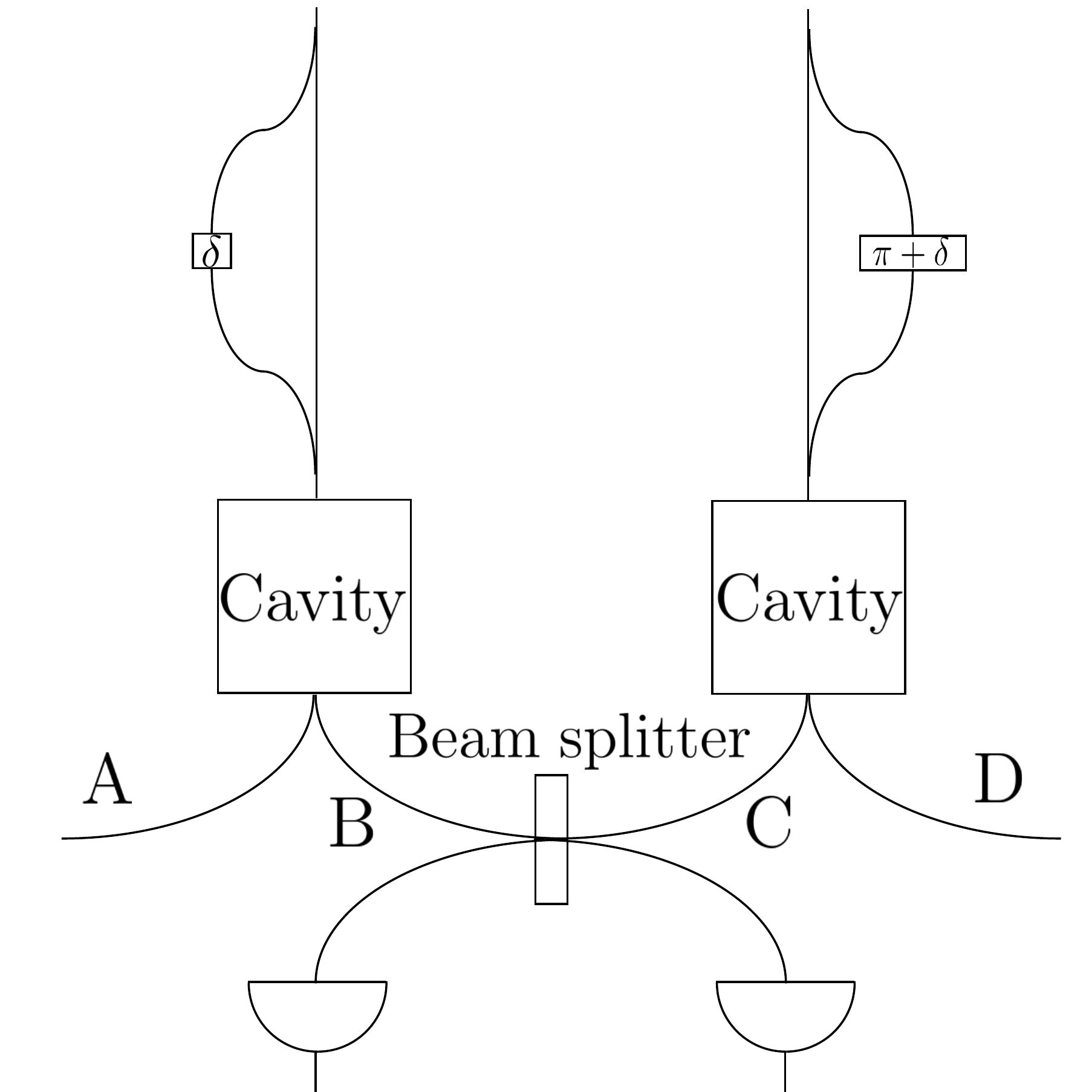}
    \caption{Time-bin encoding scheme. The pulses first go through preparation devices with the same path difference and different phase shifts. The time-bin encoded quantum states are generated by the cavities. Then Alice and Bob send photons to the middle for entanglement swapping. Photon detection in the middle heralds the Bell state between Alice and Bob. }
    \label{fig:time-bin encoding}
\end{figure}

We consider the single-photon based protocol as shown in Fig.~\ref{fig:time-bin encoding}. 
Alice and Bob produce the pump pulses by preparation devices with the same path difference and phase shifts $\delta$ and $\delta+\pi$. Due to the path difference, pump pulse states are in superposition of early and late time-bin. The pump pulse at Alice's side $\ket{\psi}_{\rm{P}_1} = \frac{1}{\sqrt{2}}(\ket{{\rm e}}+e^{i\delta} \ket{{\rm l}})$ while at Bob's side $\ket{\psi}_{\rm{P}_2} = \frac{1}{\sqrt{2}}(\ket{{\rm e}}-e^{i\delta} \ket{{\rm l}})$, where $\ket{\rm{e}}$ and $\ket{\rm{l}}$ are the `early' state and `late' state of a single photon, respectively.
Then the input pump pulses are sent to the cavities to generate M-O pairs. We assume a weak pump, then an output M-O pair can be generated from one cavity with probability $P_{11}=\text{Tr}(\hat{\rho}_{\rm m,o}\ketbra{11}{11})$, where $\hat{\rho}_{\rm m,o}$ is the Gaussian state with covariance matrix given by Eq.~\eqref{eq:CM} of the main text. 

We calculate the probability $P_{11}$ by integrating Wigner functions
\begin{align}
    \text{Tr}(\hat{\rho}_{\rm m,o}\ketbra{11}{11}) = & (2\pi)^2 \int \diff^4x_1x_2p_1p_2 W(x_1,p_1,x_2,p_2;\hat{\rho}_{\rm m,o}) \nonumber\\ 
    & \times W(x_1,p_1,x_2,p_2;\ketbra{11}{11}).
\end{align}
The Wigner function of Fock state $\ketbra{n}{n}$ is 
\begin{equation}
    W_n(x,p)=\frac{1}{\pi}(-1)^n\exp{-(x^2+p^2)}\; L_n\left(2(x^2+p^2)\right),
\end{equation}
where $L_n$ is Laguerre polynomial. Substituting this into above equation we get
\begin{align}
     &P_{11}=\text{Tr}(\hat{\rho}_{\rm m,o}\ketbra{11}{11})=\nonumber \\
     &\quad \frac{4(1+6v^2+v^4-2u w v^2-w^2-u^2+u^2 w^2)}{(1+u-v^2+w+u w)^3}.
     \label{eq:P1}
\end{align}
Similarly, we also obtain the probabilities of getting zero photon in microwave domain and one photon in optical domain 
\begin{equation}
    P_{01}={\rm{Tr}}(\hat{\rho}_{\rm m,o}\ketbra{01}{01}) = \frac{4(-1-u-v^2+w+uw)}{(1+u-v^2+w+uw)^2},
\end{equation}
one photon in microwave domain and zero photon in optical domain
\be 
P_{10}={\rm{Tr}}(\hat{\rho}_{\rm m,o}\state{10}) = \frac{4(-1+u-v^2-w+u w)}{(1+u-v^2+w+uw)^2},
\ee
and zero photon in both
\be 
P_{00}={\rm{Tr}}(\hat{\rho}_{\rm m,o}\state{00}) = \frac{4}{1+u-v^2+w+uw}.
\ee

\begin{figure}[t]
    \centering
    \includegraphics[width=0.45\textwidth]{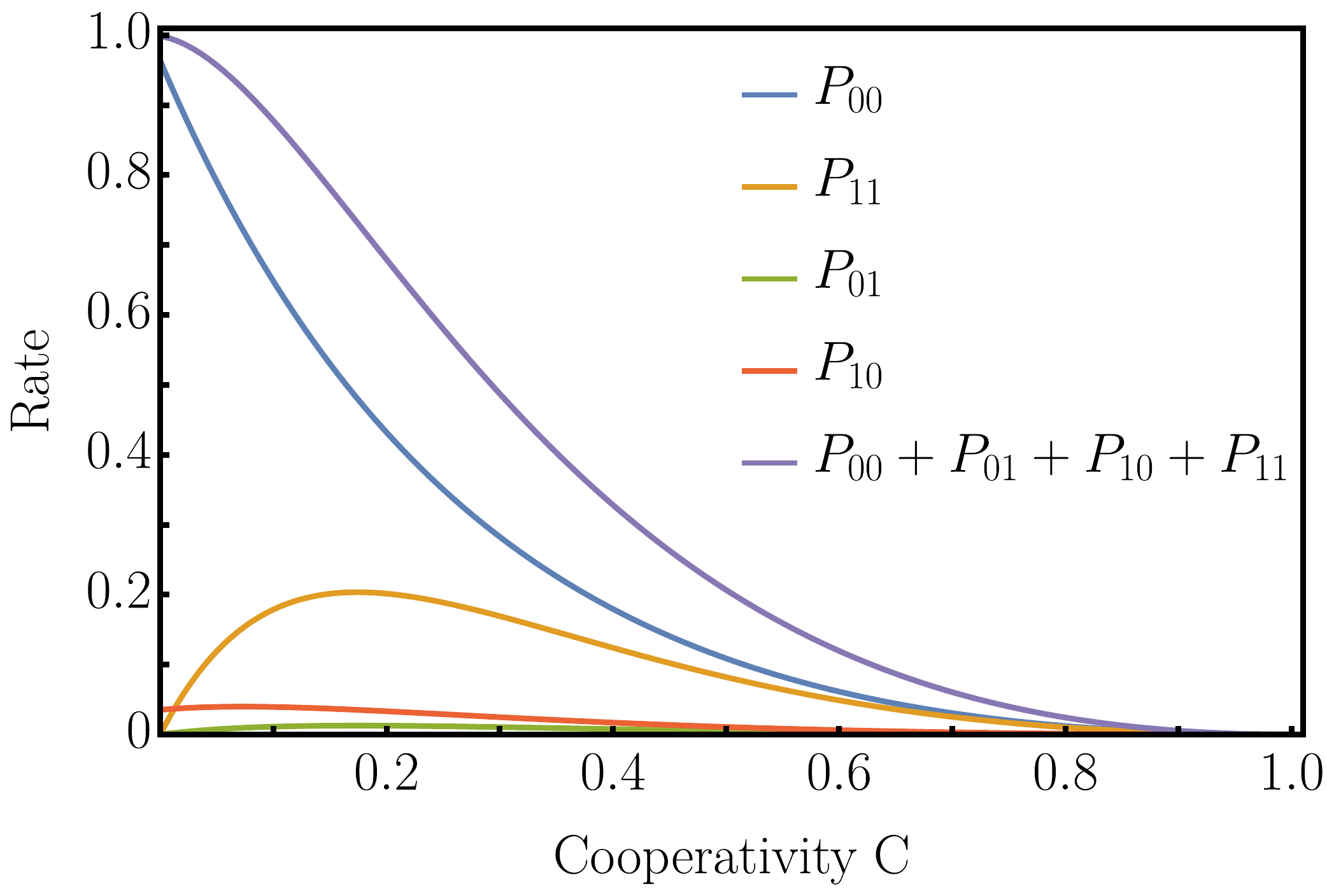}
    \caption{The probability versus the cooperativity for $\zeta_{\rm o}=0.9$, $\zeta_{\rm m}=0.95$ and $n_{{\rm in}}=0.2$.}
    \label{fig:Probability_C}
\end{figure}

Note that the weak pump region is when the cooperativity $C\ll 1$. Fig.~\ref{fig:Probability_C} shows the probabilities versus the cooperativity. 
In general, one needs to consider the generation of two or more photons in a single temporal mode in the large cooperativity region; however, such events will be beyond the consideration of this paper and will in general make the performance worse, as those events lead to states out of the single-photon Hilbert space.

Below we analyze the case where single-photon pairs are generated successfully for both Alice and Bob. At Alice's and Bob's side the M-O pairs are 
\be 
\ket{\Phi^+(\delta)}_{\rm{AB}} = \frac{1}{\sqrt{2}}(\ket{{\rm e},{\rm e}}+e^{i\delta}\ket{{\rm l},{\rm l}}),
\ee 
and 
\be 
\ket{\Phi^-(\delta)}_{\rm{CD}} = \frac{1}{\sqrt{2}}(\ket{{\rm e},{\rm e}}-e^{i\delta}\ket{{\rm l},{\rm l}}),
\ee 
respectively. The overall prepared state is in a product \cite{de_Riedmatten_2005}
\begin{align}
    &\ket{\Phi^+(\delta)}_{\rm{AB}}\otimes\ket{\Phi^-(\delta)}_{\rm{CD}} =  \frac{1}{2}\left[\ket{\Phi^+}_{\rm{BC}}\otimes\ket{\Phi^-(2\delta)}_{\rm{AD}}\right. \nonumber\\
    &\left. \quad + \ket{\Phi^-}_{\rm{BC}}\otimes \ket{\Phi^+(2\delta)}_{\rm{AD}}+\ket{\Psi^+}_{\rm{BC}} \otimes e^{i\delta}\ket{\Psi^-}_{\rm{AD}}\right. \nonumber\\
    &\left. \quad +\ket{\Psi^-}_{\rm{BC}} \otimes e^{i\delta}\ket{\Psi^+}_{\rm{AD}}\right],
\end{align}
where four Bell states are $\ket{\Phi^{\pm}(\delta)} = \frac{1}{\sqrt{2}}(\ket{{\rm e},{\rm e}} \pm e^{i\delta}\ket{{\rm l},{\rm l}})$ and $\ket{\Psi^{\pm}(\delta)}=\frac{1}{\sqrt{2}}(\ket{{\rm l},{\rm e}} \pm e^{i\delta}\ket{{\rm e},{\rm l}})$. Here $\ket{\Phi^{\pm}}$ and $\ket{\Psi^{\pm}}$ refer to the case of $\delta=0$. Bell-state measurement in the middle consists of a balanced beam splitter and two photon detectors. The detection process is as follows. To detect
\begin{align}
    \ket{\Psi^+}_{\rm{BC}} & = \frac{1}{\sqrt{2}}(\ket{{\rm e},{\rm l}}+\ket{{\rm l},{\rm e}}),\nonumber\\
    &= \frac{1}{\sqrt{2}}(\hat{a}^{(1)\dagger}_B\hat{a}^{(2)\dagger}_C+\hat{a}^{(2)\dagger}_B\hat{a}^{(1)\dagger}_C)\ket{0},
    \label{eq:Psi+}
\end{align}
where $\hat{a}^{(i)\dagger}_{B/C}$ is the creation operator for port $B/C$, time-bin $i$. The beam splitter transforms the bosonic modes by
\begin{equation}
    \begin{pmatrix}
\hat{a}_B\\
\hat{a}_C
\end{pmatrix}=
    \frac{1}{\sqrt{2}}
\begin{pmatrix}
1 & 1\\
1 & -1
\end{pmatrix}
\label{eq:beam_splitter}
\begin{pmatrix}
\hat{a}_{B^\prime}\\
\hat{a}_{C^\prime}
\end{pmatrix},
\end{equation}
where $B^\prime/C^\prime$ corresponds to the output.

Substituting Eq.~\eqref{eq:beam_splitter} into Eq.~\eqref{eq:Psi+} we get 
\begin{align}
    \ket{\Psi^+}_{BC} \rightarrow \frac{1}{\sqrt{2}}(\hat{a}^{(1)\dagger}_{B^\prime}\hat{a}^{(2)\dagger}_{B^\prime}-\hat{a}^{(1)\dagger}_{C^\prime}\hat{a}^{(2)\dagger}_{C^\prime})\ket{0}.
\end{align}
Similarly we have 
\begin{align}
    \ket{\Psi^-}_{BC} \rightarrow \frac{1}{\sqrt{2}}(\hat{a}^{(1)\dagger}_{C^\prime}\hat{a}^{(2)\dagger}_{B^\prime}-\hat{a}^{(1)\dagger}_{B^\prime}\hat{a}^{(2)\dagger}_{C^\prime})\ket{0}.
\end{align}
Detecting a single photon successively at one detector $B^\prime$ and $C^\prime$ measures $\ket{\Psi^+}_{\rm{BC}}$ and heralds $\ket{\Psi^-}_{\rm{AD}}$. Detecting a single photon at both detectors but with a time delay measures $\ket{\Psi^-}_{\rm{BC}}$ and heralds $\ket{\Psi^+}_{\rm{AD}}$. The above two are success events in generating a Bell pair.

The other two Bell states transform as
\begin{align}
    \ket{\Phi^\pm} \rightarrow  \frac{1}{2\sqrt{2}}& \left[(\hat{a}^{(1)\dagger}_{B^\prime}\hat{a}^{(1)\dagger}_{B^\prime}-\hat{a}^{(1)\dagger}_{C^\prime}\hat{a}^{(1)\dagger}_{C^\prime})\right. \nonumber\\
    &\left. \quad \pm (\hat{a}^{(2)\dagger}_{B^\prime}\hat{a}^{(2)\dagger}_{B^\prime}-\hat{a}^{(2)\dagger}_{C^\prime}\hat{a}^{(2)\dagger}_{C^\prime})\right]\ket{0}.
\end{align}
In these cases, we will detect two photons at one of the detectors $B^\prime$ and $C^\prime$. Because there is no time difference between the detections, we can not distinguish these two cases.

When BC is projected to either $\ket{\Psi^+}_{\rm{BC}}$ or $\ket{\Psi^-}_{\rm{BC}}$, then AD ends up with $e^{i\delta}\ket{\Phi^-}_{\rm{AD}}$ or $e^{i\delta}\ket{\Phi^+}_{\rm{AD}}$. The probability of success is limited to $50\%$ with perfect detection.

Now we consider the other events that are considered as success, while entanglement is not generated. When both B and C have photons, it will trigger success events.

The first case is when A gets zero photon, while B,C,D all get one photon. This happens with probability $P_{01}P_{11}$. 
In this case, the states before swap are
\ba
\ket{\Phi^+(\delta)}_{\rm{AB}} = \frac{1}{\sqrt{2}}(\ket{0,{\rm e}}+e^{i\delta}\ket{0,{\rm l}}),
\\ 
\ket{\Phi^-(\delta)}_{\rm{CD}} = \frac{1}{\sqrt{2}}(\ket{{\rm e},{\rm e}}-e^{i\delta}\ket{{\rm l},{\rm l}}).
\ea
Replacing $\ket{0}$ in the above analysis, after we have success detection, the state between AD ends in
\begin{align}
\ket{\Psi^{\pm}}_{{\rm{AD}}}
&=\frac{1}{\sqrt{2}}(\ket{0,{\rm e}} \pm \ket{0,{\rm l}}),
\\
&=\ket{0}_A \otimes \frac{1}{\sqrt{2}}(\ket{{\rm e}} \pm \ket{{\rm l}})_D.
\end{align}

Similarly, in the second case, when D gets zero photon, and A,B,C all have one photon, with probability also $P_{01}P_{11}$, we have the final state
\begin{align}
\ket{\Psi^{\pm}}_{{\rm{AD}}}
&=\frac{1}{\sqrt{2}}(\ket{{\rm e}} \pm \ket{{\rm l}})_A \otimes \ket{0}_D.
\end{align}

The final case is when A and D have zero photon, while B and C have one photon, which happens with probability $P_{01}^2$. In this case, we will have 
\begin{align} 
\ket{\Psi^{\pm}}_{{\rm{AD}}}&=\ket{0,0},
\end{align} 
as a tensor product of vacuum.

So the `success' event happens with probability 
\be 
P_{\rm success}=(P_{11}^2+2P_{01}P_{11}+P_{01}^2)/2.
\ee 
The overall state conditioned on success is equivalent to the following state,
\begin{widetext}
\begin{align}
\hat{\rho}_{AD}& =\frac{1}{P_{11}^2+2P_{01}P_{11}+P_{01}^2}\left[P_{11}^2\state{\rm Bell}
+P_{01}P_{11} \state{0}\otimes \state{+}+P_{01}P_{11}\state{+}\otimes\state{0}+P_{01}^2 \state{00}
\right],\\
& \equiv \lambda_1 \state{\rm Bell}
+\lambda_2 \state{0}\otimes \state{+}+\lambda_2 \state{+}\otimes\state{0}+\lambda_3 \state{00},
\label{eq:rho_timebin}
\end{align}
\end{widetext}
where we defined $\lambda_1, \lambda_2,\lambda_3$ implicitly.
The entanglement generation rate is therefore
\be 
R(\hat{\rho}_{\rm AD}) = E(\hat{\rho}_{AD}) P_{\rm success}/2,
\ee 
where we divide by 2 since we use the channel twice for time-bin encoding and the $E()$ can be the RCI or EOF for entanglement measure.

The RCI of $\hat{\rho}_{{\rm{AD}}}$ is evaluated as the following. First the entropy 
\be
S(\hat{\rho}_{{\rm{AD}}})=-\lambda_1 \log_2(\lambda_1)-2\lambda_2 \log_2(\lambda_2)-\lambda_3 \log_2(\lambda_3).
\ee
The reduced state
\begin{align}
    \hat{\rho}_{{\rm{A}}}& = {\rm{Tr}_D}(\hat{\rho}_{{\rm{AD}}}), 
    \nonumber\\
    & =\left(\frac{\lambda_1}{2}+\frac{\lambda_2}{2}\right)\state{{\rm{l}}}+\left(\frac{\lambda_1}{2}+\frac{\lambda_2}{2}\right)\state{{\rm{e}}} \nonumber\\
    &\quad + \frac{1}{2}\lambda_2 \ketbra{{\rm{e}}}{{\rm{l}}}+\frac{1}{2}\lambda_2 \ketbra{{\rm{l}}}{{\rm{e}}}+(\lambda_2+\lambda_3)\state{0}.
\end{align}
The three eigenvalues of $\hat{\rho}_{{\rm{A}}}$ are
\begin{align*}
    &\lambda_1^\prime=\frac{1}{2}\lambda_1,\\
    & \lambda_2^\prime=\frac{1}{2}(\lambda_1+2\lambda_2),\\
    &\lambda_3^\prime=\lambda_2+\lambda_3.
\end{align*}
Therefore the RCI is 
\begin{align}
    & S\left(\hat{\rho}_A\right) - S\left(\hat{\rho}_{AD}\right)
    =-\sum_{i=1}^3\lambda_i^\prime \log_2(\lambda_i^\prime)-S\left(\hat{\rho}_{AD}\right).
\end{align}

However, it is hard to evaluate EoF of $\hat{\rho}_{\rm AD}$ and thus we consider its upper bound by the fidelity to Bell pair
\be 
F_{\rm S-P}=\frac{P_{11}^2}{P_{11}^2+2P_{01}P_{11}+P_{01}^2}.
\ee
following Eq.~\eqref{eq:eof}
\begin{equation}
    E_f(\hat{\rho}_{AD}) \le \frac{P_{11}^2}{P_{11}^2+2P_{01}P_{11}+P_{01}^2} S(\ket{\rm Bell}) = F_{\rm S-P},
\end{equation}
as the pure states mixture in Eq.~\eqref{eq:rho_timebin} are orthorgonal.

\section{Optical transmission loss}
\label{app:optical_loss}
\begin{figure}[t]
  \centering
    \includegraphics[width=0.35\textwidth]{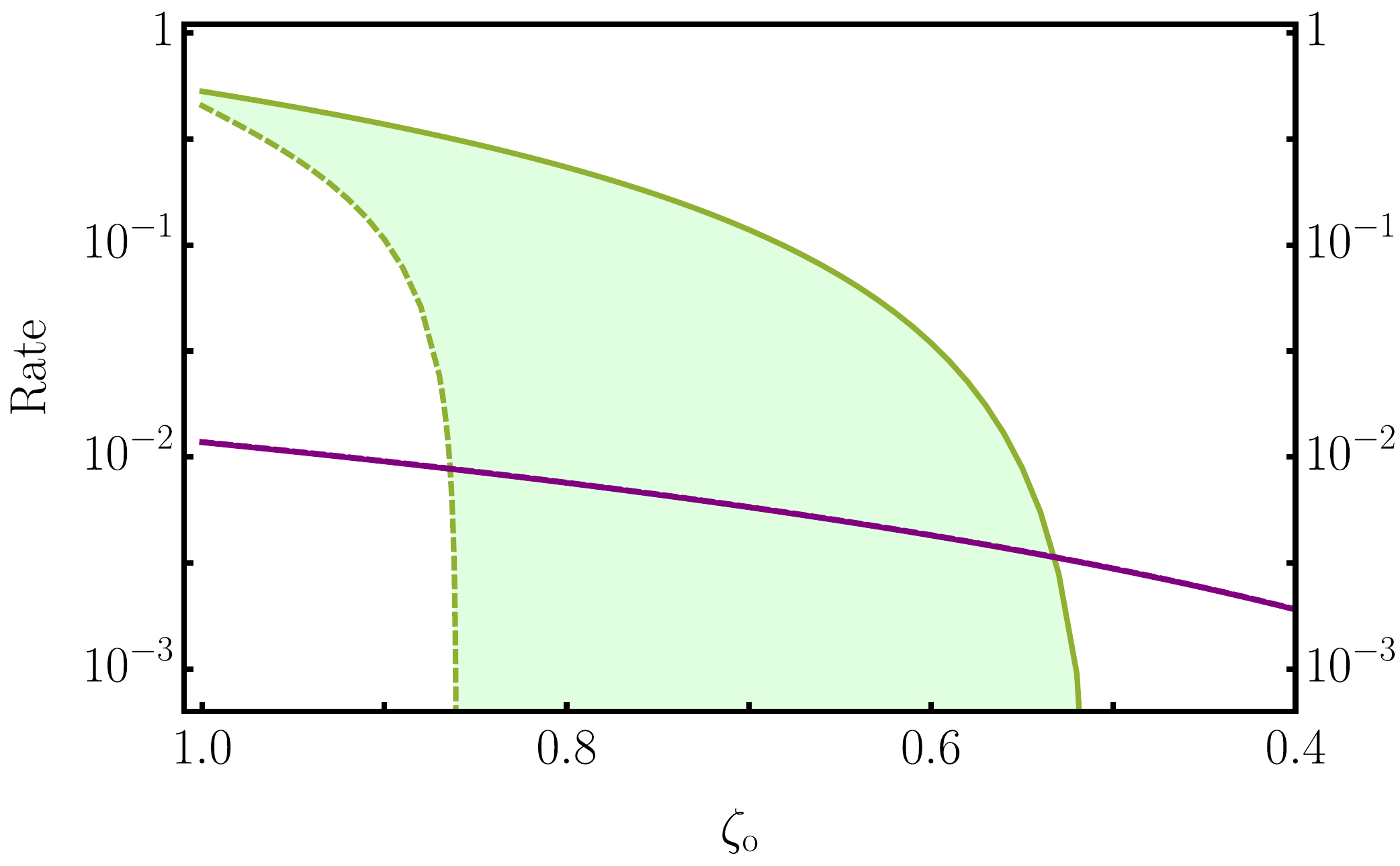}
    \caption{Rate versus $\zeta_{\rm o}$ when $\zeta_{\rm m}=0.992$, $n_{{\rm {in}}}=0.2$ and $C=0.1$. Solid and dashed green curves represent lower and upper bounds for rate of CV entanglement swap. Purple line show rate of time-bin entanglement swap. 
    \label{rate_loss}
    }
\end{figure}

We consider identical pure loss channels with transmissivity $\eta$ before the beam splitter. The channel is described by
\be
\hat{a}\to \sqrt{\eta} \hat{a}+\sqrt{1-\eta}\hat{e},
\ee
where $\hat{a}$ and $\hat{e}$ represent the optical mode and the environmental vacuum separately. 
In the CV entanglement swap case, the parameters under the loss are transformed as
\begin{subequations}
\begin{align}
& u \to  u=1+\frac{8\zeta_{\rm m} [{{C}}+ n_{\rm in}(1-\zeta_{\rm m})]}{(1-{{C}})^2},\\
& v \to \sqrt{\eta} v = \frac{4\sqrt{\zeta_{\rm o}\eta \zeta_{\rm m} {{C}}}[1+{{C}}+2 n_{\rm in}(1-\zeta_{\rm m})]}{(1-{{C}})^2},\\
& w \to (1-\eta) + \eta w = 1+\frac{8{{C}}\zeta_{\rm o}\eta\left[1+n_{\rm in}\left(1-\zeta_{\rm m}\right)\right]}{(1-{{C}})^2}.
\end{align}
\end{subequations}
One can easily check that, transmitting the noisy entangled CV modes through a pure loss channel of transmissivity $\eta$ is equivalent to the state with $\zeta_{\rm {o}}\to \eta \zeta_{\rm{o}}$, via Eqs.~\eqref{uvw} in the main text. When $\eta\le 0.5$ (equivalent to $\zeta_{\rm o}\le 0.5$), the CV entanglement swap enabled rate is zero, as the pure loss channel below half transmissivity has zero rate.
In Fig.~\ref{rate_loss}, we see that our approach allows an advantage when $\zeta_{\rm o} \gtrsim 0.5$, which corresponds to 15 kilometers of state-of-the-art fiber link to the center swap node.

In the time-bin entanglement case, the analysis is similar, simply replacing $\zeta_{\rm {o}}\to \eta \zeta_{\rm{o}}$ in previous expressions of Sec.~\ref{App:timebin}. Note that one cannot simply multiply the success probability by $\eta^2$, as two or more photons can lead to a single photon after loss. 


\section{Solving the direct swap}
\label{appendix:direct_swap}

\begin{figure}[t]
    \centering
    \includegraphics[width=0.45\textwidth]{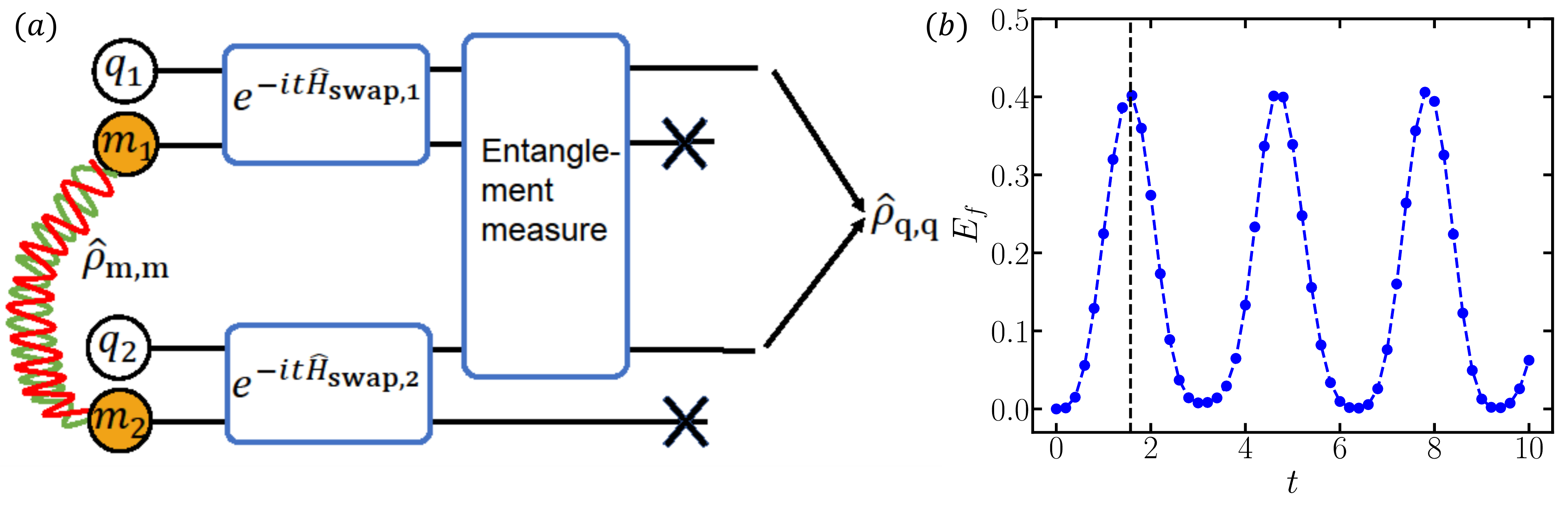}
    \caption{(a) Schematic of the direct swap approach. (b) Entanglement of formation $E_f$ between qubits versus evolution time $t$. Vertical dashed line indicates the time with maximum entanglement at $t=\pi/2$.}
    \label{fig:direct_swap_scheme}
\end{figure}

In this section, we provide more details on the direct swap approach, for entanglement distillation on the hybrid CV-DV platform. For a given pair of noisy entangled microwave modes in state $\hat{\rho}_{\rm m,m}$, the composite system is prepared same as it is in the hybrid LVQC approach where two qubits are in state $\ket{0}_{\rm q}$, shown in the top and bottom halves of Fig.~\ref{fig:direct_swap_scheme}(a). The CV-DV systems on both sides are evolved separately by the same unitary $\exp(-it\hat{H}_{{\rm swap},\ell})$ where $\ell=1,2$ stands for the two sides. Through the evolution, we monitor the entanglement (i.e. EoF) between two qubits, as shown in Fig.~\ref{fig:direct_swap_scheme}(b). We stop the evolution and discard the modes when the two qubits are maximally entangled at the time $t=(2n+1)\pi/2, n\in\mathbb{N}$.


In the direct swap approach, we choose the final two-qubit state to maximize the entanglement, whose fidelity to $\ket{\Psi^+}$ is not guaranteed to be the maximal. To maximize the fidelity, we allow an arbitrary local unitary $\hat{U}$ on one of the qubit. Note that the Bell state $\ket{\Psi^+}$ is invariant under $\hat{U}\otimes \hat{U}^\star$, so all local unitary can be absorbed into a single qubit. We parameterize the single qubit unitary by three angles and numerically maximize the fidelity to obtain the results.

\section{Qubit distllation protocols}
\label{appendix:qubit_distill}

\begin{figure}[t]
    \centering
    \includegraphics[width=0.5\textwidth]{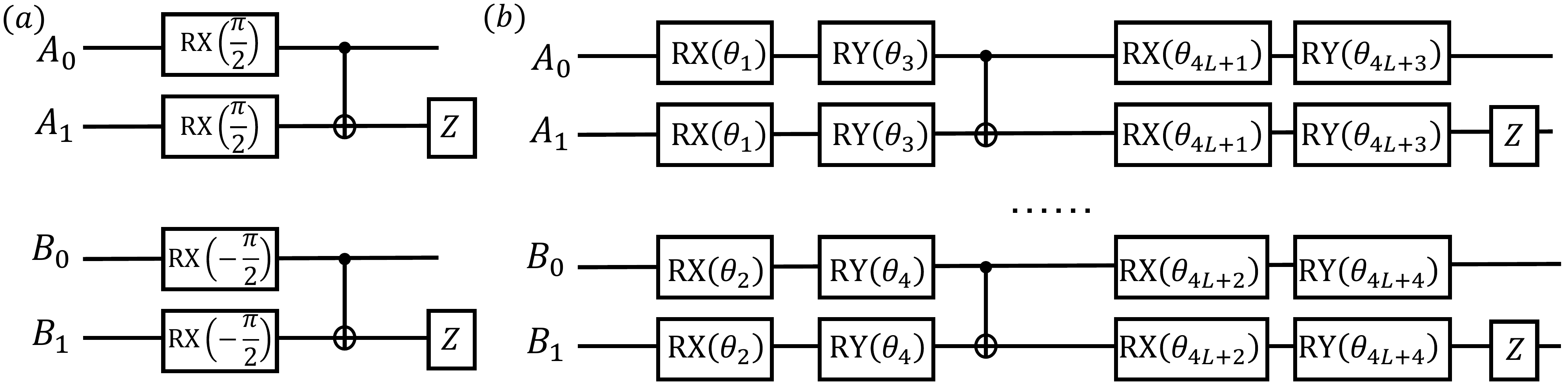}
    \caption{Schematic of (a) DEJMPS protocol and (b) $L$-layer DV LVQC.}
    \label{fig:dv_protocols}
\end{figure}

As explained in the main text, to obtain states with further lower infidelity, we consider distillation protocols on two-copy of the output qubits produced from hybrid LVQCs or direct swap approach. Various entanglement distillation protocols on DV systems have been proposed, including BBPSSW~\cite{bennett1996purification}, DEJMPS~\cite{deutsch1996} and LOCCNet~\cite{zhao2021practical}. 
We mainly focus on the DEJMPS protocol which is proved to be optimal for bell-diagonal states with rank up to three, and a DEJMPS-inspired DV LVQC, shown in Fig.~\ref{fig:dv_protocols}(a)(b) separately. For both protocols, the two-qubit system $A_0, B_0$ and $A_1, B_1$ are separately initialized with identical two-qubit mixed state produced from hybrid distillation approach. Through the circuit, qubits $A_1, B_1$ are measured in the computational basis where only $\ket{00}$ or $\ket{11}$ are considered as success.
The DV LVQC is trained in a similar way to the way we train the hybrid LVQC described in our main text and details can be found in Fig.~\ref{fig:distillation_detail}(c)(d).


\section{Numerical details}
\label{app:numerical_details}

In this section, we provide more details on the numerical simulation of the hybrid LVQC approach. For each choice of hyperparameters $\lambda, F_c$, we start with a batch of $500$ instances with random initialization.
The quantum circuits are implemented in PyTorch~\cite{paszke2019pytorch} and optimized by Adam~\cite{kingma2014adam} with the learning rate ${\rm lr}=0.001$ in $20000$ steps. Unless further specified, in this section we utilize hybrid LVQCs with $D = 10$ gate blocks and DV LVQCs with $L=6$ layers. In the distillation, we choose post-selection on photon number below $D/2$ as success.

\subsection{Details of Fig.~\ref{fig:distillation_performance} of the main text}

\begin{figure}[t]
    \centering
    \includegraphics[width=0.35\textwidth]{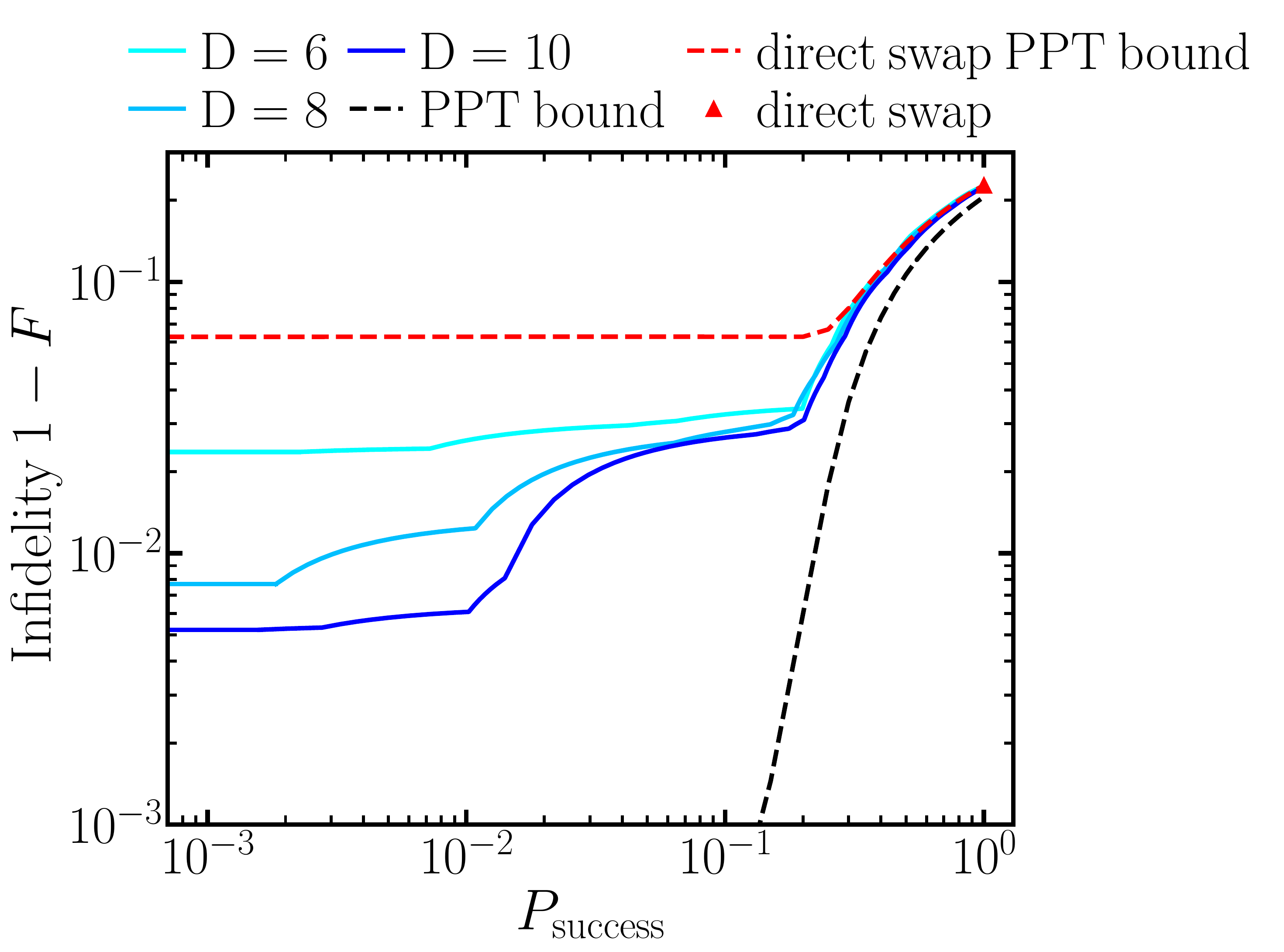}
    \caption{Infidelity versus success probability of one copy noisy entangled microwave modes with $C=0.1, n_{\rm in}=0.2, \zeta_{\rm m}=0.992, \zeta_{\rm o}=0.99$ by utilizing hybrid LVQC with $D$ ECD blocks.
        \label{fig:distillation_depth}
        }
\end{figure}

To begin with, we analyze the effect of circuit depth of the hybrid LVQC. We consider the same setting as Fig.~\ref{fig:distillation_performance}(a) of the main paper, but choose different LVQC depth $D$ and the results are in Fig.~\ref{fig:distillation_depth}. As we see, in the low infidelity region, the performance does improve as the depth increases. Indeed, a larger depth $D\ge 10$ can potentially further improve the performance, and further enlarge the advantage. 

\begin{figure*}
    \centering
    \includegraphics[width=0.65\textwidth]{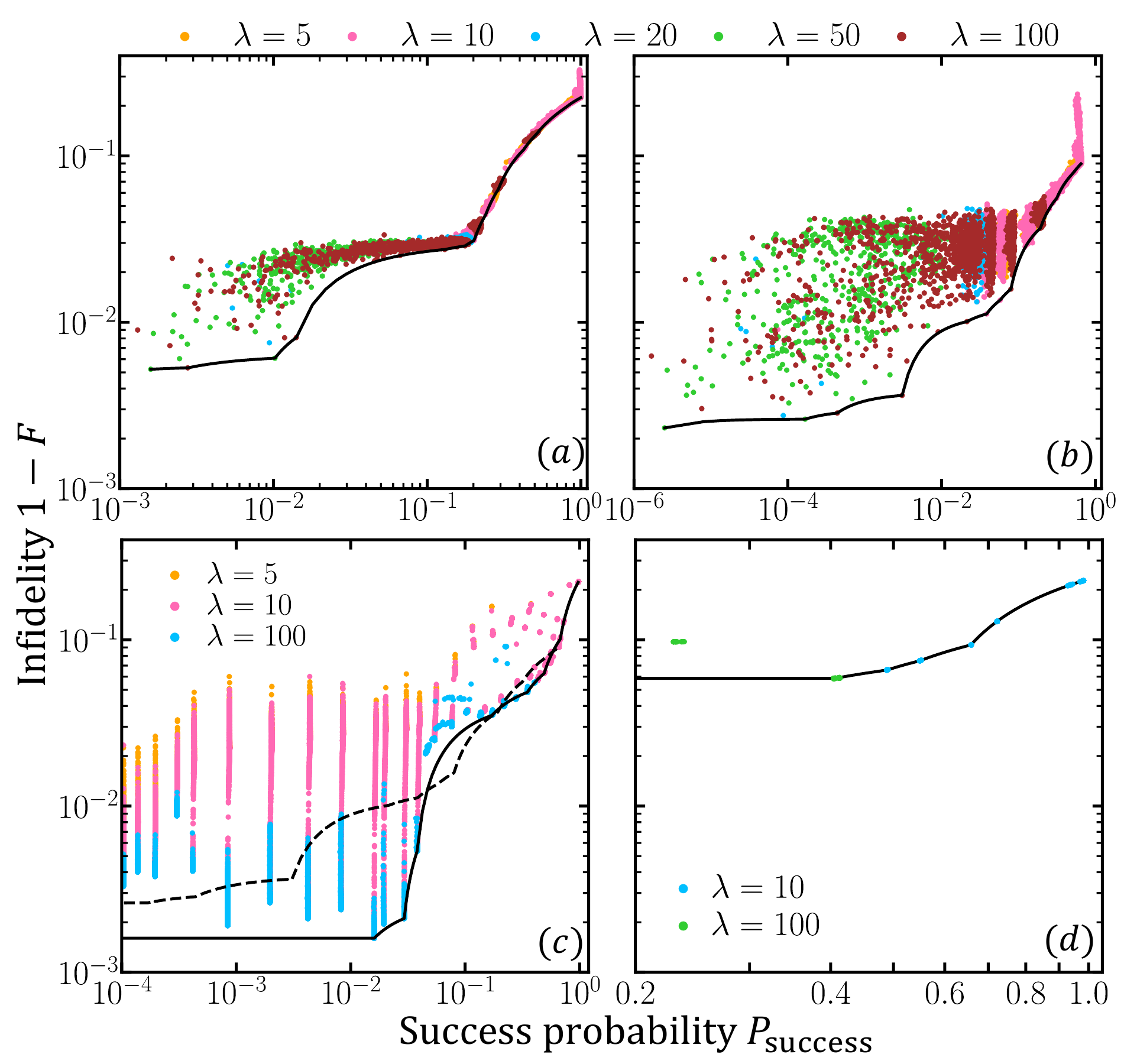}
    \caption{Numerical simulation details of infidelity versus success probability with (a) hybrid LVQC; (b) hybrid LVQC+DEJMPS; (c) hybrid LVQC+DV LVQC; (d) direct swap+DV LVQC. Scatter dots with different color represent training results with different choice of hyperparameters $\lambda, F_c$ and random initializations. Black solid curves show the smallest infidelity for given success probability for each approach with interpolation. Black dashed curve in (c) represents the smallest infidelity for hybrid LVQC+DEJMPS (same as black solid curve in (b)).
        \label{fig:distillation_detail}
    }
    \centering
    \includegraphics[width=0.85\textwidth]{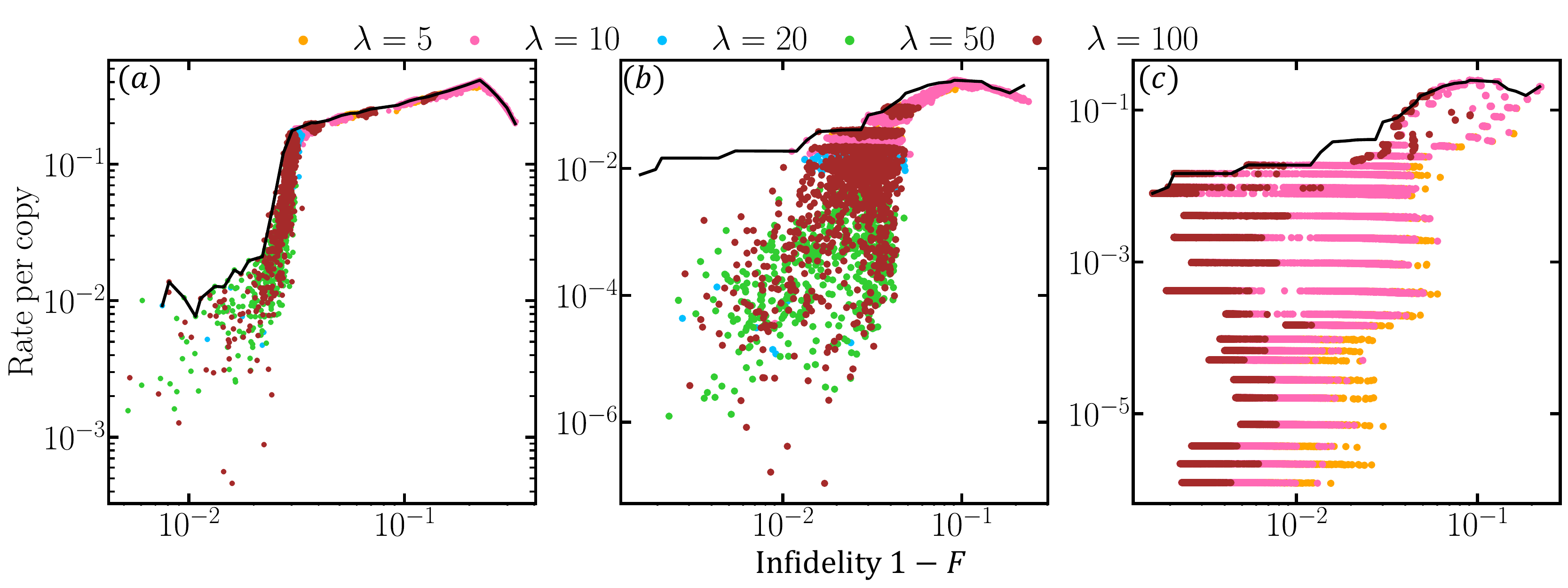}
    \caption{Numerical simulation details of distillation rate $R_{E_f}$ versus infidelity for (a) one-copy hybrid LVQC, (b), (c) two-copy hybrid LVQC+DEJMPS and hybrid LVQC+DV LVQC. Scatter dots with different color represent training results with different choice of hyperparameters $\lambda, F_c$ and random initializations. Black solid curves show the highest rate given infidelity for (a) one-copy and (b)(c) two copy.
    \label{fig:distillation_detail_2}
    }
\end{figure*}

Now we present more details on the training results utilizing hybrid LVQCs. In Fig.~\ref{fig:distillation_detail}(a)(b), we show the results of the one-copy case, and the two-copy DEJMPS case. To fulfill the whole range of success probability $P_{\rm success}\in[0,1]$, we adopt different combinations of hyperparameters $\lambda, F_c$. 
To obtain the best performance from hybrid LVQCs in the two-copy scenario, we extract the instances with highest fidelity within each small bin of success probability and apply interpolation (black solid curves) between them. The details of the interpolation will be introduced in the next paragraph. In Fig.~\ref{fig:distillation_detail}(c), we show training results with DV LVQC (see Fig.~\ref{fig:dv_protocols}(b) for the circuit design) on those best output qubit pairs in (a) with various combinations of $\lambda, F_c$. Note that there is advantages from DV LVQCs in terms of infidelity compared to DEJMPS, and thus the combination of the DEJMPS and DV LVQCs provides the best DV approach results for two-copy case, shown in Fig.~\ref{fig:distillation_performance}(b) in the main text. In Fig.~\ref{fig:distillation_detail}(d), we apply the same DV LVQC approach on the two-qubit state generated by direct swap. 
To summarize, compared to the DEJMPS protocol, the DV LVQC approach can provide better performance with smaller infidelity at the expense of more quantum gates.

To obtain the best performance of variational circuits for each given success probability, we utilize interpolation and extrapolation~\cite{rozpkedek2018optimizing}. For any two protocol with success probability $p_1, p_2$ and fidelity $F_1, F_2$, we consider the probabilistic mixing of the protocols with probability $r, 1-r\in[0,1]$, then the success probability and fidelity for the protocol mixture are
\begin{subequations}
\begin{align}
    P_{\rm success,int} &= rp_1 + (1-r)p_2,\\
    F_{\rm int} &= \frac{rp_1F_1 + (1-r)p_2 F_2}{P_{\rm success,int}}.
    \label{eq:interpolation}
\end{align}
\end{subequations}
Similarly, one can always extrapolate a protocol to a trivial strategy that always output a product of $\ket{0}_q$ state, with an identity success probability but $1/2$ fidelity. The protocol with extrapolation has success probability and fidelity as
\begin{subequations}
\begin{align}
    P_{\rm success,ext} &= rp_1 + (1-r),\\
    F_{\rm ext} &= \frac{rp_1F_1 + (1-r)/2}{P_{\rm success,ext}}.
    \label{eq:extrapolation}
\end{align}
\end{subequations}

\subsection{Details of Fig.~\ref{fig:distillation_rate} of the main text}

In Fig.~\ref{fig:distillation_detail_2}, we provide the details of the rate versus infidelity plot of Fig.~\ref{fig:distillation_rate} in the main text. We present the data points from different training, under various hyperparameters indicated by the different coloring. Subplot (a) presents the single-copy case, while (b) and (c) present the two-copy case, with the DEJMPS and DV LVQC separately. The black solid lines in (b)(c) are the best performance of the results in both (b) and (c), which are eventually presented in Fig.~\ref{fig:distillation_rate}(b) of the main text.

%

\end{document}